\documentclass{article}

\usepackage{arxiv}
\usepackage[utf8]{inputenc} 
\usepackage[T1]{fontenc}    
\usepackage{hyperref}       
\usepackage{url}            
\usepackage{booktabs}       
\usepackage{amsmath}
\usepackage{amsfonts}
\usepackage{amssymb}
\usepackage{nicefrac}       
\usepackage{microtype}      
\usepackage{lipsum}
\usepackage{graphicx}
\usepackage{caption}
\usepackage{placeins}
\usepackage{tabularray}
\usepackage{booktabs} 
\usepackage{subcaption}
\usepackage{longtable}
\usepackage{algorithm}
\usepackage{algorithmic}

\graphicspath{ {./images/} }

\usepackage{fancyhdr}

\fancypagestyle{plain}{%
    \fancyhf{} 
    \fancyfoot[R]{\thepage} 
}

\title{Exploiting Precision Mapping and Component-Specific Feature Enhancement for Breast Cancer Segmentation and Identification}

\author{
 Pandiyaraju V \\
  School of Computer Science and Engineering\\
  Vellore Institute of Technology, Chennai\\
  Tamil Nadu, India\\
  \texttt{pandiyaraju.v@vit.ac.in} \\
   \And
 Shravan Venkatraman \\
  School of Computer Science and Engineering\\
  Vellore Institute of Technology, Chennai\\
  Tamil Nadu, India \\
  \texttt{shravan.venkatraman18@gmail.com} \\
  \And
  Saraswathi D \\
  School of Computer Science and Engineering\\
  Vellore Institute of Technology, Chennai\\
  Tamil Nadu, India \\
  \texttt{saraswathi.d@vit.ac.in} \\
  \And
 Pavan Kumar S \\
  School of Computer Science and Engineering\\
  Vellore Institute of Technology, Chennai\\
  Tamil Nadu, India \\
  \texttt{s.pavankumar2003@gmail.com} \\
\And
 Santhosh Malarvannan \\
  School of Computer Science and Engineering\\
  Vellore Institute of Technology, Chennai\\
  Tamil Nadu, India \\
  \texttt{sandy5501m@gmail.com} \\
  \And
 Kannan A \\
  Department of Information Science and Technology\\
  College of Engineering, Guindy, Anna University, Chennai\\
  Tamil Nadu, India \\
  \texttt{akannan123@gmail.com} \\
}

\begin{document}
\maketitle
\begin{abstract}
Breast cancer is one of the leading causes of death globally, and thus there is an urgent need for early and accurate diagnostic techniques. Although ultrasound imaging is a widely used technique for breast cancer screening, it faces challenges such as poor boundary delineation caused by variations in tumor morphology and reduced diagnostic accuracy due to inconsistent image quality. To address these challenges, we propose novel Deep Learning (DL) frameworks for breast lesion segmentation and classification. We introduce a precision mapping mechanism (PMM) for a precision mapping and attention-driven LinkNet (PMAD-LinkNet) segmentation framework that dynamically adapts spatial mappings through morphological variation analysis, enabling precise pixel-level refinement of tumor boundaries. Subsequently, we introduce a component-specific feature enhancement module (CSFEM) for a component-specific feature-enhanced classifier (CSFEC-Net). Through a multi-level attention approach, the CSFEM magnifies distinguishing features of benign, malignant, and normal tissues. The proposed frameworks are evaluated against existing literature and a diverse set of state-of-the-art Convolutional Neural Network (CNN) architectures. The obtained results show that our segmentation model achieves an accuracy of 98.1\%, an IoU of 96.9\%, and a Dice Coefficient of 97.2\%. For the classification model, an accuracy of 99.2\% is achieved with F1-score, precision, and recall values of 99.1\%, 99.3\%, and 99.1\%, respectively.

\end{abstract}


\section{Introduction}

Breast cancer is one of the most common cancers among women worldwide, causing about 570,000 deaths in 2015 alone. Every year, more than 1.5 million women, or 25\% of all female cancer diagnoses, are diagnosed with breast cancer worldwide \cite{stewart2014world}\cite{whoBreastCancer}. Breast tumors often begin as ductal hyperproliferation and can become benign tumors or metastatic carcinomas when stimulated by various carcinogenic agents. The tumor microenvironment, such as stromal effects and macrophages, contributes to the development and progression of breast cancer \cite{sun2017risk}.

Early detection of breast carcinoma increases the chances of successful treatment. Thus, there is a need for the implementation of effective procedures for the early detection of breast cancer signs \cite{tarique2015fourier}. Mammography, ultrasound, and thermography are some of the primary imaging techniques used in screening and diagnosing breast cancer \cite{acsDiagnosis}\cite{sadoughi2018artificial}. With over 75\% of the tumors being responsive to hormones, breast cancer is mostly a postmenopausal disease. Their incidence rates are at their peak ages between 35-39 years and then stabilize after the age of 80 years with age and female sex as major risk factors. This hormone dependence combines with environmental and genetic factors in determining the incidence and natural history of the disease \cite{benson2009early}. Thus, accurate segmentation and classification of breast cancer are important for proper treatment planning and positive patient outcomes.

The traditional methods rely a lot on manual interpretation that consumes a lot of time and has room for errors. Advancements in technology have transformed the provision of healthcare \cite{alz}. High processing power, primarily from GPUs, enables the creation of deep neural networks with multiple layers, allowing for the extraction of formerly unachievable features. CNNs have made a profound impact on image processing and understanding, especially in the areas of medical image segmentation, classification, and analysis \cite{litjens2017survey}\cite{rashed2019deep}. DL models can process vast amounts of medical imaging data and detect subtle abnormalities that might elude human observers. Accurate tumor segmentation and classification enhances oncologists' capacity to make decisions about whether a tumor is malignant or not. Typically, these methods require professional annotation and pathology reports to make this assessment \cite{ramesh2022segmentation}, which consumes a lot of human effort. DL provides an efficient and promising solution for the automation of these procedures \cite{skin}. They can learn complicated patterns and features from ultrasounds and mamograms, which has the potential to improve classification accuracy and efficiency.

We introduce, in this paper, a novel Precision Mapping and Attention-Driven LinkNet framework, with an integrated Precision Mapping Mechanism (PMM) module. PMM dynamically calibrates pixel-level mappings to achieve accurate boundary delineation, compensating for morphological variations and inconsistencies in medical images. A specially designed attention-driven architecture, leveraging a pre-trained encoder backbone, optimizes feature extraction. The decoder uses a coupled spatial-channel attention mechanism to dynamically refine spatial features with enhanced semantic context for accurate tumor segmentation. We also propose a Component-Specific Feature Enhancement Module (CSFEM) to develop a Component-Specific Feature-Enhanced classifier - CSFEC-Net. Selective amplification of discriminative patterns allows the CSFEM to dynamically learn and amplify tissue-specific characteristics that improve feature representation, and enhance inter-class separability. Segmentation results were validated by employing the Dice coefficient, IoU score, and a combination of focal loss with Jaccard loss. On the other hand, classification is evaluated using the following evaluation metrics: recall, F1-score, precision, and accuracy.

Our proposed contributions are:
\begin{itemize}
\item \textbf{Precision Mapping Mechanism (PMM):} A novel boundary refinement module for a PMAD-LinkNet segmentation framework that integrates an adaptive mapping strategy with pixel-level precision to localize tumor edges with high accuracy. This is done by dynamically calibrating spatial mappings according to morphological variations and compensating for inconsistencies in image quality through a learned refinement process.
\item \textbf{Component-Specific Feature Enhancement Module (CSFEM):} A dedicated feature augmentation module for CSFEC-Net that uses dynamic multi-attention mechanisms to selectively amplify discriminative traits of benign, malignant, and normal tissues. Using a multi-scale attention pipeline, CSFEM enhances inter-class separability and intra-class consistency and thus can effectively address the problem of overlapping feature spaces in heterogeneous medical imagery.
\end{itemize}

The organization of this paper is as follows: Section 2 reviews the literature on breast cancer segmentation and classification; Section 3 describes the proposed approach; Section 4 presents experimental results; Section 5 concludes and outlines future research directions.

\section{Related Works}
\label{sec:headings}
Osareh et al. \cite{osareh2010machine} utilized the K-nearest neighbors (KNN), Support Vector Machine (SVM), and Probabilistic Neural Network (PNN) classification models to perform the classification of tumor regions. The methodology was employed on two different publicly available datasets where one of the datasets was composed of Fine Needle Aspirates of the Breast Lumps (FNAB) with 457 negative samples and 235 positive samples while the other dataset was composed of 295 gene microarrays with 115 good-prognosis class and 180 poor-prognosis class data. To support the classifier, feature extraction and selection methodologies were utilized. Feature extraction techniques like Principal Component Analysis (PCA), optimized with auto-covariance coefficients of feature vectors, were employed to reduce high-dimensional features into low-dimensional ones. Feature selection includes two different approaches such as the Relief algorithm for filter approach where the features are selected using a pre-processing step and no bias of the induction algorithms is considered unlike the wrapper approach namely the proposed Sequential forward selection (SFS) technique where a feature set composed of 15 sonographic features are obtained. The results underwent ranking using a feature ranking method that employed Signal-to-Noise Ratio (SNR) to identify crucial features. The evaluation involved wrapper approach estimates assessed through a leave-one-out cross-validation procedure, focusing on overall accuracy, Sensitivity, Specificity, and Matthews Correlation Coefficient (MCC).

Li et al. \cite{li2019classification} introduced a novel patches screening method that included the extraction of multi-size and discriminative patches from histology images involving tissue-level and cell-level features. Firstly, patches of dimensions 512x512 and 128x128 are generated from the input data. This is followed by the utilization of two ResNet50s where one of the models is fed with patches of dimensions 128x128 while the other inputs patches of dimensions 512x512 which extract tissue-level and cell-level features respectively. A finetuning approach is adopted to train the ResNet50 models this is followed by a screening of patches by aggregating them into different clusters based on their phenotype. For speeding up the process, the patch size is reduced to obtain 1024 features followed by PCA to reduce the number of features to 200. This is followed by the k mean clustering process. A ResNet50 fine-tuned with 128x128 size patches is employed to select the clusters. Subsequently, the P-norm pooling feature method is applied to extract the final features of the image, followed by the use of a Support Vector Machine to classify input images into four distinct classes: Normal, Benign, In situ carcinoma, or Invasive carcinoma. Zheng et al. \cite{zheng2020deep} introduced a DL-assisted Efficient Adaboost Algorithm (DLA-EABA) where the Convolutional Neural Network is trained with extensive data so that high precision can be achieved. A stacked autoencoder is utilized for generating a deep convolutional neural network and the encoder and decoder sections contain multiple non-linear transformations which are taken from the combined depictions of actual data which is taken as input. An efficient Adaboost algorithm is utilized to train the classifiers which estimate the positive value for threshold and parity and is done by reviewing all the potential mixtures of both values, The deep CNN contains Long Short-Term Memory (LSTM) with logistic activation function as conventional artificial neurons. This is followed by Softmax Regression for classifying the images with the help of features extracted.

Lotter et al. \cite{lotter2021robust} introduced a robust breast tumor classification model for mammography images which utilizes bounding box annotations and is extended to digital breast tomosynthesis images to be able to identify the tumor region in the image. The CNN first trains to classify if lesions are present in the cropped image patches. Subsequently, using the entire image as input, the CNN initializes the backbone of the detection-based model. This model outputs the entire image with a bounding box, providing a classification score. The model's performance is then evaluated by comparing its ability to identify the tumor region with Breast Imaging Reporting and Data System Standard (BI-RADS) scores of 1 and 1 considered as negative interpretations and index and pre-index cancer exams. Saber et al. \cite{saber2021novel} employed transfer learning methodology on five different models: ResNet50, VGG19, Inception V3, Inception-V2, and VGG16. Feature extraction involved freezing the trained parameters from the source task except for the last three layers, which were then transferred to the target task. The images were preprocessed using different methods such as Median Filter, Histogram Equalization, Morphological Analysis, Segmentation, and Image Resizing. The dataset is split into an 80-20 ratio and Augmentation is applied to the training dataset where the images are rotated and flipped. The newly trained layers are combined with the existing pre-trained layers and the features are extracted using these models. Classification is done by feeding the extracted features from the transfer learning models into a Support Vector Machine classifier and Softmax classifiers that are fine-tuned using the Stochastic Gradient Descent method with momentum (SGDM). The gradient’s high-velocity dimensions are reduced due to SGDM jittering and the past gradients with momentum are reduced to saddle point.  

Cho et al. \cite{cho2022deep} proposed a Breast Tumor Ensemble Classification Network (BTEC-Net) which utilizes an improved DenseNet121 and ResNet101 as base classifiers where each of the four blocks is connected to the Squeeze and Excitation Block and Global Average Pooling layer. Next, the feature map sizes are aligned using a fully connected layer and integrated along the channel dimension. The combined feature map is then fed into a feature-level fusion module to perform binary classification. Once the classification is done, segmentation is carried out by utilizing the proposed Residual Feature Selection UNet model (RFS-UNet) which is an encoder-decoder network and are connected with the layer positions of the same feature map size using skip-connections. In the encoder part, there are five encoders and each one includes a convolutional layer, an RFS module, a residual convolutional block and a max-pooling layer. The model also consists of five decoders in which each decoder contains a convolutional layer and an RFS module, a transpose convolutional layer and a Residual Block as well. The skip connections are equipped with a spatial attention module which takes as input the output of the transposed convolution and the output of the RFS from the encoder, and produces concatenation of outputs to the output of the same transposed convolution layer. The last layer in the segmentation process uses a sigmoid activation function which retrieves the output which is the segmented tumor region. Dayong Wang et al.\cite{wang2016deep} presented a new approach for detecting metastatic breast cancer in whole-slide images of sentinel lymph node biopsies and won the first place in the ISBI 2016 grand challenge. Their system performed satisfactorily with an AUC of 0.925 for whole-slide image classification and 0.7051 as a tumor localization score that surpassed an independent pathologist’s review. Combining predictions of the DL system and diagnostics of pathologists, a significant decrease of the error rate was obtained.

Abdelrahman Sayed Sayed et al. \cite{hamed2020deep} proposed a novel low-cost structure of a 3-RRR Planar Parallel Manipulator (PPM) in their research – attempting to address the problem of formulating kinematic constraint equations for drive manipulators with complex nonlinear dependence. They performed direct and inverse kinematics techniques using screw theory and established a Neuro-Fuzzy Inference System (NFIS) model optimized with Particle Swarm Optimization (PSO) and Genetic Algorithm (GA) for end-effector position prediction. Proposed PPM structure was emphasized, including kinematic modeling and preliminary PPM virtual tests within ADAMS, followed by verification prototype manufacture. Results demonstrated PSO as a more successful tuning approach for the NFIS model than GA, which provided better consistency with the experimental PPM data and thus the strategy holds potential for further robot enhancement and performance improvement opportunities from additional optimization and control measures. A thorough review by Luuk Balkenende et al. \cite{balkenende2022application} described an integrated application of deep learning techniques in imaging breast cancer. Their paper demonstrates a variety of applications in digital mammography, ultrasound colonography, and magnetic resonance imaging and introduces some applications such as lesion classification, segmentation, and therapy response prediction. The research is also highlighted by findings on the autonomous diagnosis of breast cancer metastasis with convolutional neural networks on whole-body scintigraphy scans and the investigation into helping clinicians diagnose axillary lymph node metastases using a 3D CNN model on PET/CT images. They mention the necessity to conduct large-scale trials and address ethical considerations in order to realize the full potential of deep learning in clinical breast cancer imaging.

A novel approach to the detection of breast cancer in screening mammograms based on deep learning was presented by Shen et al. \cite{shen2019deep}. This "end-to-end" algorithm is claimed to use efficiently training sets of various levels of annotation to perform really well by comparison with previous methods. Across the independent test sets collected from different mammography platforms, the method reaches per-image AUCs ranging from 0.88 to 0.98, coupled with sensitivities in the range of 86.1\%-86.7\%. Most important is that it has demonstrated cross-platform transferability, which requires only a minimal amount of additional data for fine-tuning. The results underlined the promise of deep learning in transforming breast cancer screening into more accurate and efficient diagnostic tools and thus into clinical applications. A novel approach for breast cancer diagnosis and prognosis employing a class structure-based Deep CNN was established by Han et al \cite{han2017breast}. In automated multi-class classification from histopathological images, their CSDCNN addresses the caveat and does an impressive average accuracy of 93.2\%. By establishing a hierarchical feature representation with distance constraints in feature space, this methodology serves as a unique avenue to minor differences between the various classes of breast cancer to which comparative experiment results are significantly favorable for the CSDCNN compared to other methods. It also poises itself as one of the most valuable clinical decision aids in breast cancer management. This study has provided an important impetus in the direction of automated breast cancer classification and reliable diagnoses aid to clinicians.

 Sizilio et al. \cite{sizilio2012fuzzy} proposed a fuzzy logic approach to pre-diagnosing breast cancer from FNA analysis. The method was introduced in the context of the global burden of breast cancer and also a wide range of accuracy in FNA-based diagnosis, from 65 to 98 percent. Hence, this method increases the reliability of computing intelligence. The study made use of the Wisconsin Diagnostic Breast Cancer Data (WDBC) and followed fuzzification, rule-base creation, inference processing, and defuzzification. Validation was based on cross-validation and expert reviews. The method achieved a sensitivity of 98.95\% and specificity of 85.43\%, implying a very reliable malignant detection but a very poor benign detection. This technique has a good potential for improving diagnostic accuracy in breast cancer. The application of K-Nearest Neighbours (KNN) for breast cancer diagnosis with the Wisconsin-Madison Breast Cancer dataset was explored by Sarkar et al. \cite{sarkar2000application}. KNN was relatively straightforward and efficient in its non-parametric classifier use in this respect. This research showed that KNN increased the classification performance with an improvement of 1.17\% over the best-known result for the dataset. KNN is simple, effective for smaller training sets, and it does not require retraining of trained models when new data is available. However, limitations also arise as it incurs huge storage overhead for larger datasets and high computation requirements to calculate distances between test and training data. The study mentions that there are faster KNN variants, such as k-d trees, which have performed excellently in script and speech recognition tasks. KNN has much more potential in regard to diagnostic applications, even though one algorithm is not suitable for all problems in diagnostics. This research states not only the promise of KNN in improving the accuracy of diagnostics but also addresses the challenges of storage and computational efficiency. 

Foster et al. \cite{foster2014machine} provided a critical overview of the role of ML in biomedical engineering, paying attention to the extension of SVMs beyond the use of this statistical tool. Their study revealed the intrinsic challenges in constructing SVM-based clinically validated diagnostic systems, pointing out problems, such as overfitting and the need for thorough validation protocols. Unlike disease-specific studies, their study was designed to evaluate and enhance existing ML models for a broader biomedical application. The commentary is a cautionary view for researchers, reviewers, and readers in emphasizing the complexities and pitfalls that may accompany classifier development. It supports an integrated approach where validation of classifiers is part of the experimental process. This work puts great emphasis on the fact that, in biomedical research, clinical validity should be ascertained for the diagnosis tools developed by ML. Wei et al. \cite{wei2009microcalcification} introduced an innovative approach which improves the classification performance of the proposed diagnostic system based on the classification of breast cancer images by microcalcifications, which is developed through a combination of CBIR and ML. The combination of CBIR improves the retrieval of similar mammogram cases for enhanced SVM classifier performance. Utilizing recovered case information with local proximity, the adaptive SVM achieved significantly better classification accuracy compared with the baseline model at 78\% to 82\%, as measured in the area under the ROC curve. The technique could make for a good source of support for radiologists while delivering an enhanced "second opinion" tool. Still, the study underlines problems with small datasets that may well influence generalization ability. These results suggest that assistive classification approaches using CBIR may improve the accuracy in the diagnosis of breast cancers, but further validation, possibly with larger clinical data sets, will be required to confirm such efficacy and practical applicability in actual clinical settings.

\section{Proposed Work}

\subsection{Methodology}

\begin{figure}[h!]
  \centering
  \includegraphics[width=0.7\textwidth]{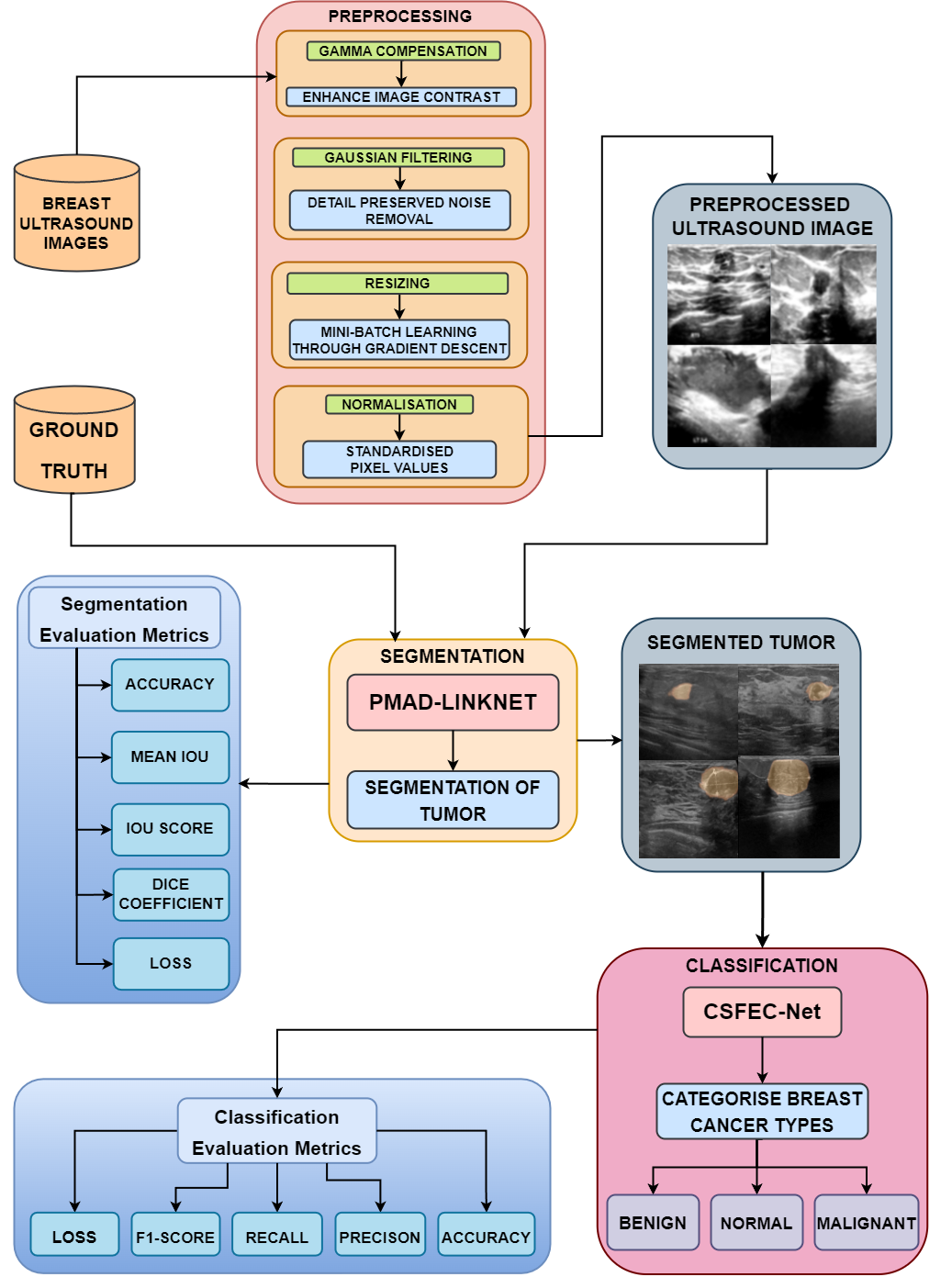}
  \caption{Overall Workflow Diagram of Proposed Work}
  \label{fig:overallArch}
\end{figure}

Augmentation is applied initially to the ultrasound images dealing with class imbalance. In the subsequent stage, a series of preprocessing steps comprising gamma correction, gaussian filtering, resizing, and normalizing are used on these images. Pixel values are known to contain non-linearity in ultrasound images primarily in the regions of very high or low intensity. Gamma correction would be apt to counter this effect, therefore producing images more accurate and appealing to the human eye. An effective technique for noise reduction and edge detail preservation in ultrasound images is the application of Gaussian filtering. This effectively reduces noise while preserving edge details. To preserve consistency throughout the dataset and facilitate batch processing, resizing is done to ensure the images fed into the proposed DL model have the same dimensions. Normalizing the image pixels to scale within a specific range enhances the quality of activation functions' ability to capture the non-linearities in the data. Here, the images have been scaled to fall within the range (0, 1). The preprocessed images are then fed to the proposed Spatial-Channel Attention LinkNet Framework with InceptionResNet backbone for segmenting the tumor region. The segmented tumor maps are then fed to the proposed CSFEC-Net classifier to classify the segmented mass as benign, normal, or malignant. The overall workflow of this proposed work has been presented in Figure \ref{fig:overallArch}.

\subsection{Dataset Description}

\begin{figure}[h!]
  \centering
  \includegraphics[width=0.7\textwidth]{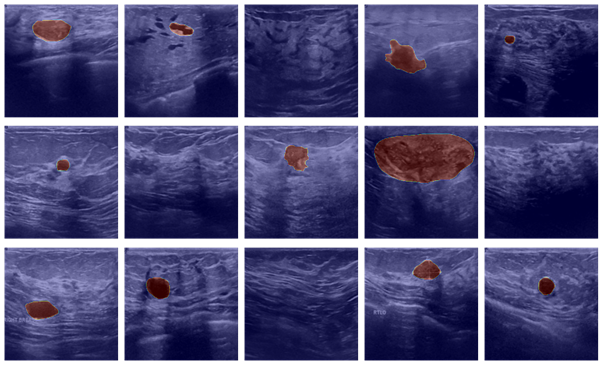}
  \caption{Samples of Breast Ultrasound Images and Masks (overlap) from the Dataset}
  \label{fig:DatasetExploration}
\end{figure}

The data utilized in this work was obtained from the Breast Ultrasound Images Dataset \cite{kaggle_breast_ultrasound} made available by Arya Shah on Kaggle. It contains a total of 780 ultrasound images along with their corresponding segmented ground truth masks, split into three categories – benign, malignant, and normal. Figure \ref{fig:DatasetExploration} showcases a sample of ultrasound images from the dataset overlapped with their corresponding segmentation maps.

\begin{figure}[h!]
    \centering
    \begin{minipage}[b]{0.48\textwidth}
        \centering
        \includegraphics[width=\textwidth]{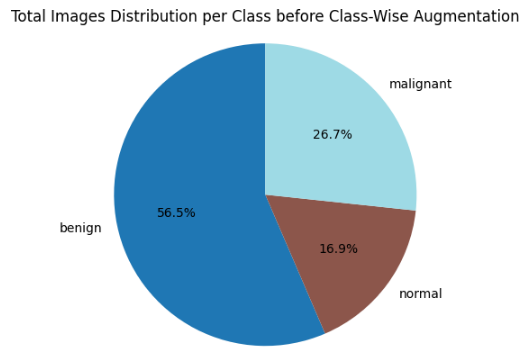} 
        \caption{Training and Validation Dice Coefficient Curves of Proposed Localization Framework}
        \label{fig:beforeAug}
    \end{minipage}
    \hfill
    \begin{minipage}[b]{0.48\textwidth}
        \centering
        \includegraphics[width=0.925\textwidth]{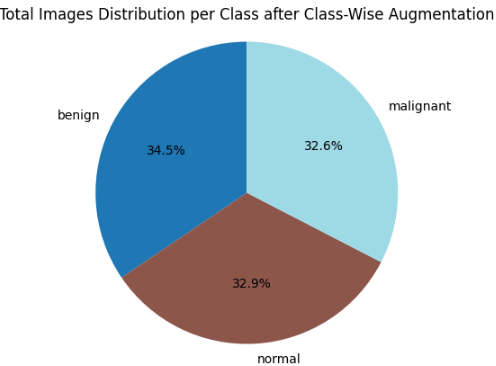} 
        \caption{Training and Validation IoU Score Curves of Proposed Localization Framework}
        \label{fig:afterAug}
    \end{minipage}
\end{figure}

The dataset exhibits a significant class imbalance, with benign samples contributing to 56.5\% of the data, while malignant and normal samples covered only 26.7\% and 16.9\% respectively. The distribution of ultrasound images exhibiting this class imbalance has been represented graphically in Figure \ref{fig:beforeAug}. To mitigate this imbalance and avoid bias during the training of segmentation and classification models, augmentation techniques are utilized. Specifically, random crop, random rotation, random zoom, random shear, and random exposure methods were applied to augment the images belonging to the 'normal' and 'malignant' classes. 

The rationale behind this augmentation approach is to level the data count of the 'normal' and 'malignant' classes, thereby aligning them more closely with the larger 'benign' class. By augmenting the training data for the 'normal' and 'malignant' classes, it is hoped that the effects of class imbalance are to be mitigated and enable the models to learn effectively from all classes. This approach ensures that the segmentation and classification models are trained on a more balanced dataset, thus improving their ability to accurately segment, identify, and characterize breast tumors across different classes. This augmentation resulted in a well-balanced data distribution of each category, which has been represented in Figure \ref{fig:afterAug}.

\subsection{Preprocessing}
\begin{figure}[h!]
  \centering
  \includegraphics[width=0.8\textwidth]{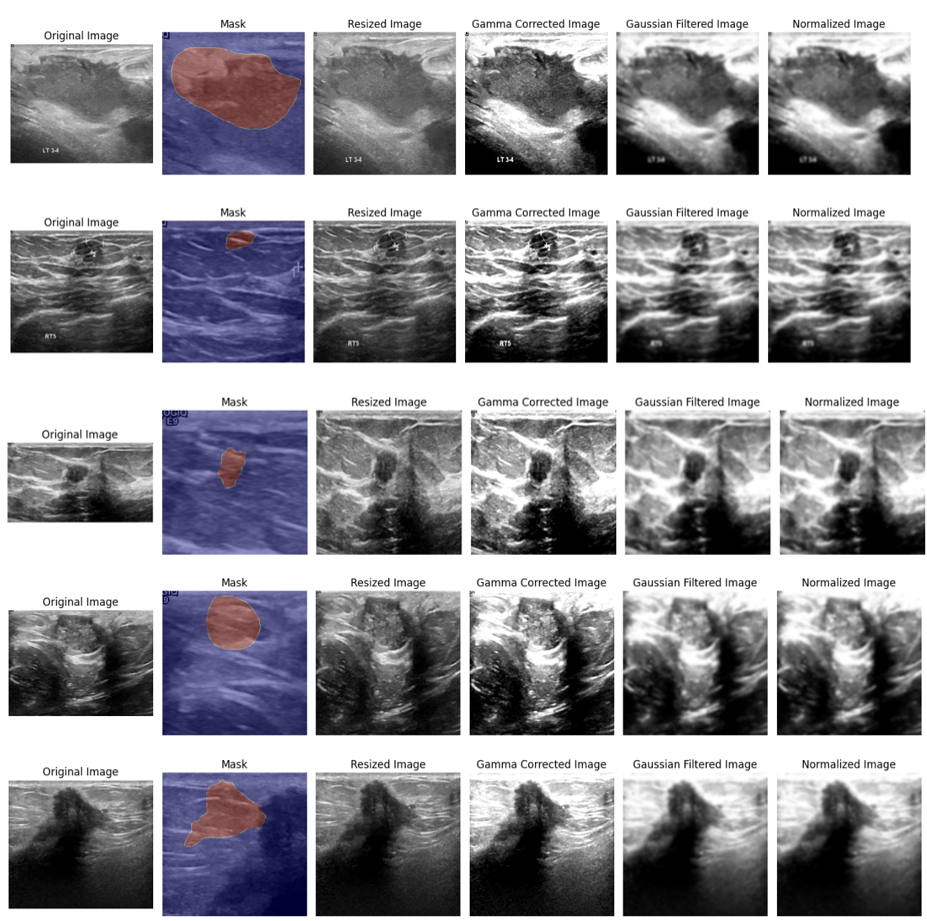}
  \caption{Breast Ultrasound Images Observed After Each Preprocessing Step}
  \label{fig:PrepOps}
\end{figure}

Following augmentation, the images were preprocessed using a pipeline, composed of four stages – gamma correction, gaussian filtering, resizing, and image normalization.  The output images obtained after each preprocessing step of breast ultrasound image preprocessing are shown in Figure \ref{fig:PrepOps}.

\subsubsection{Gamma Correction}

Gamma correction is the first preprocessing step in our ultrasound image preprocessing pipeline. It has a very important role in enhancing the visibility of key anatomical structures and subtle details in the images. Gamma correction enhances the boundary delineation of tumors and improves the visibility of tumor features by adjusting the brightness and contrast of the image. This is especially crucial for the step in breast cancer tumor segmentation since correct visualization of tumor margins helps to properly define the tumors and the regions around them for better analysis. In this, gamma correction maps the input pixel intensity, \( I_{\text{in}} \), into the output pixel intensity, \( I_{\text{out}} \), as determined by the value of the gamma \( \gamma \) given in Equation~\ref{eq:gamma_correction}.

\begin{equation} \label{eq:gamma_correction}
\underbrace{I_{out}}_{\text{Output Pixel Intensity}} = \underbrace{I_{in}}_{\text{Input Pixel Intensity}}^\gamma
\end{equation}

\subsubsection{Gaussian Filtering}

After gamma correction, we apply Gaussian filtering in order to reduce speckle noise, a prevalent artefact in ultrasound images which can obscure the boundaries of a tumor and prevent its proper segmentation. By selectively smoothing out noise while preserving necessary details, Gaussian filtering improves the clarity of tumor features and enhances the accuracy of our segmentation algorithms. This step is critical for segmenting and classifying tumor in breast cancer: since it reduces noise artifacts as well as increases the fidelity of tumor delineation to the extent that more accurate and reliable segmentation results are obtained. In summary, Gaussian filtering ensures that our images are cleaner and, by this, more amenable to subsequent segmentation and classification steps, thus allowing us to detect and characterize the tumors properly. The mathematical representation of the Gaussian filtering process is as follows (Equation \ref{eq:GF}), where \( I_{\text{in}}(x,y) \) represents the input pixel intensity at coordinates \( (x, y) \), \( I_{\text{out}}(x,y) \) represents the output pixel intensity after Gaussian filtering, and \( G(i,j) \) represents the Gaussian kernel applied to the neighborhood of the pixel. The size of the Gaussian kernel is represented by \( N \).

\begin{equation} \label{eq:GF}
\underbrace{I_{out}(x,y)}_{\text{OutputImage}} = \sum_{i=-\frac{N}{2}}^{\frac{N}{2}} \sum_{j=-\frac{N}{2}}^{\frac{N}{2}} \underbrace{I_{in}(x+i,y+j)}_{\text{InputImage}} \cdot \underbrace{G(i,j)}_{\text{GaussianKernel}}
\end{equation}

\subsubsection{Ultrasound Image Resizing}

Once the images have been gamma corrected and Gaussian filtered, we resize to standardize image dimensions, making them compatible with our segmentation and classification algorithms. Standardized image dimensions \( (x', y') \) are important for ensuring consistency and comparability across different datasets and analysis pipelines. By resizing, we change the input image to an output image: \( I_{\text{in}}\left(\frac{x}{r_x}, \frac{y}{r_y}\right) \)\(\rightarrow\)\( I_{\text{out}}(x', y') \) with scaling factors \( r_x \) and \( r_y \) along the \( x \)- and \( y \)-axes, respectively. This way, we can get a uniform framework for analysis that makes the processing pipeline less complex with reduced computational complexity, as in Equation~\ref{eq:resizing}.

\begin{equation} \label{eq:resizing}
\underbrace{I_{out}(x',y')}_{\text{OutputImage}} = \underbrace{I_{in}\left(\frac{x}{r_x}, \frac{y}{r_y}\right)}_{\text{InputImage}}
\end{equation}

\subsubsection{Pixel Normalization}

 After gamma correction and then application of the Gaussian filtering to the images, normalization will be used to map pixel values within the normalized image to an appropriate, consistent range - usually to lie between 0 and 1. This procedure normalizes the intensity of in-input pixels to ensure equivalence between different images as required when preparing images to train neural networks. By transforming the input image \( I_{\text{in}} \) to the output image \( I_{\text{out}} \), where

\begin{equation}\label{eq:normalization}
\underbrace{I_{out}}_{\text{Output Image}} = \frac{\underbrace{I_{in}}_{\text{Input Image}} - \underbrace{\min(I_{in})}_{\text{Minimum Of Input Image}}}{\underbrace{\max(I_{in})}_{\text{Maximum Of Input Image}} - \underbrace{\min(I_{in})}_{\text{Minimum Of Input Image}}}
\end{equation}

we eliminate variations in intensity that may arise due to differences in acquisition parameters or imaging conditions. The algorithmic workflow of our preprocessing pipeline is presented in Algorithm \ref{alg:preprocessing}. 

\begin{algorithm}
\caption{Breast Ultrasound Image Preprocessing}
\label{alg:preprocessing}
\begin{algorithmic}[1]
\REQUIRE 2D breast ultrasound images
\ENSURE Preprocessed breast ultrasound images
\STATE \textbf{Function} \texttt{preprocess\_image}():
\STATE $height, width, \_ \leftarrow \text{get\_dimensions}(\text{input})$
\STATE $\gamma \leftarrow \text{gamma\_correction\_factor}$
\FOR{$h \leftarrow 0$ \TO $height$}
    \FOR{$w \leftarrow 0$ \TO $width$}
        \STATE $gamma\_corrected\_image(h, w) \leftarrow \text{input}(h, w)^\gamma$
    \ENDFOR
\ENDFOR
\FOR{$h \leftarrow 0$ \TO $height$}
    \FOR{$w \leftarrow 0$ \TO $width$}
        \STATE $gaussian\_filtered\_image(h, w) \leftarrow$
        \STATE \hspace{1cm} $\displaystyle\sum_{i=-h/2}^{h/2} \sum_{j=-w/2}^{w/2} gamma\_corrected\_image(h+i, w+j) \cdot G(i, j)$
        \STATE $G(i, j) = \dfrac{1}{2\pi\sigma^2} \exp\left(-\dfrac{i^2 + j^2}{2\sigma^2}\right)$
    \ENDFOR
\ENDFOR
\STATE $r_h, r_w \leftarrow \text{resizing factors for image of size } (height, width)$
\STATE $h', w' \leftarrow \dfrac{height}{r_h},\ \dfrac{width}{r_w}$
\STATE $resized\_image \leftarrow gaussian\_filtered\_image[r_h{:}h',\ r_w{:}w']$
\FOR{$h \leftarrow 0$ \TO $h'$}
    \FOR{$w \leftarrow 0$ \TO $w'$}
        \STATE $normalised\_image(h, w) \leftarrow \dfrac{resized\_image(h, w) - \min(resized\_image)}{\max(resized\_image) - \min(resized\_image)}$
    \ENDFOR
\ENDFOR
\STATE $preprocessed\_image \leftarrow normalised\_image$
\RETURN $preprocessed\_image$
\end{algorithmic}
\end{algorithm}

\subsection{Precision Mapping and Attention-Driven LinkNet (PMAD-LinkNet) Framework for Breast Lesion Segmentation}

This section presents the proposed PMAD-LinkNet framework for breast cancer tumor segmentation. It takes preprocessing breast ultrasound images and their corresponding ground truth masks as input and produces the predicted segmentation map as output.

\begin{figure}[h!] 
  \centering
  \includegraphics[width=0.8\textwidth]{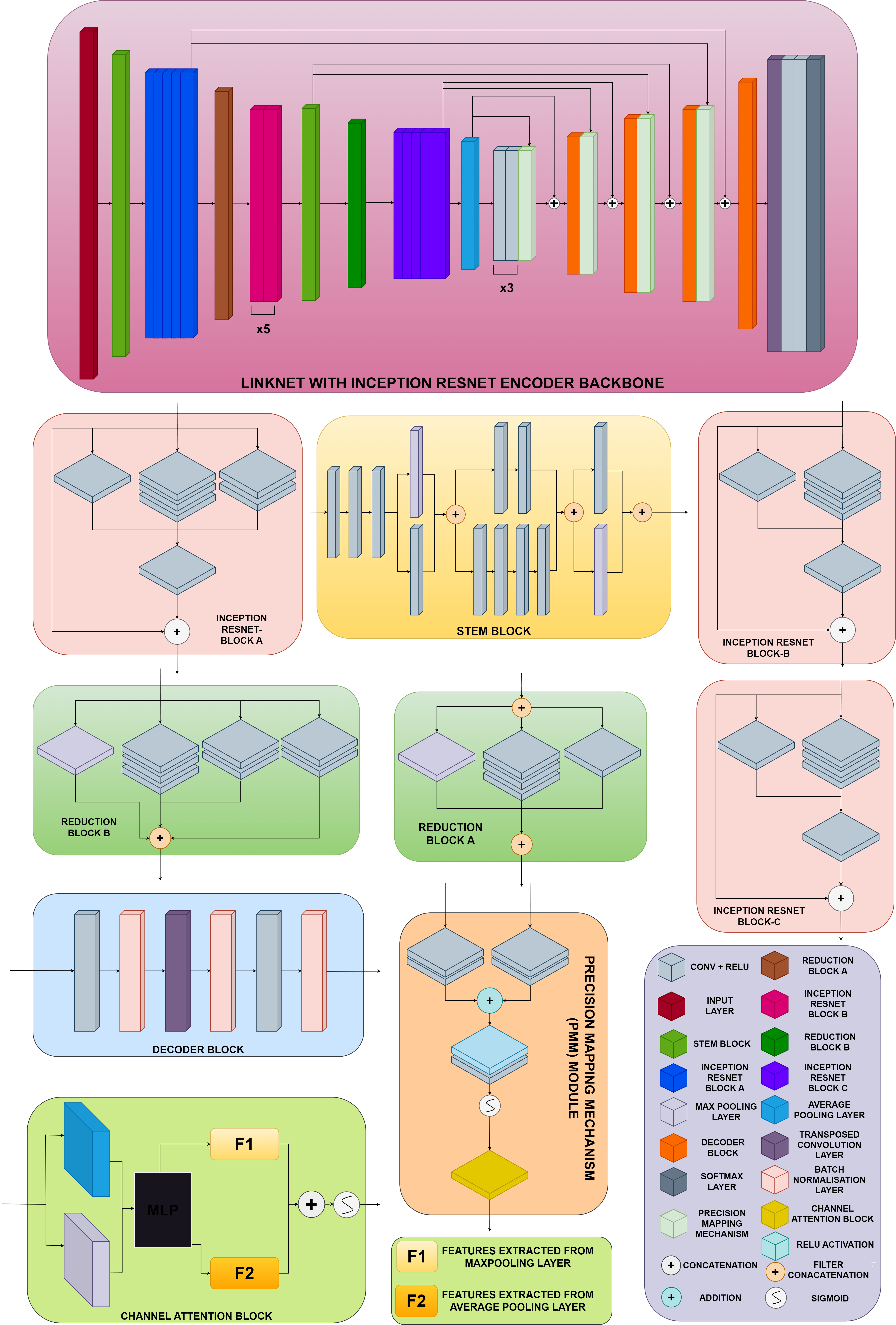}
  
  \caption{PMAD-LinkNet Layer Architecture for Breast Cancer Segmentation}

  \label{fig:segArch}
\end{figure}

The LinkNet architecture \cite{chaurasia2017linknet} is a neural network model, used for semantic segmentation especially in the field of biomedical imaging. The encoder of the proposed framework is built using an InceptionResNet backbone \cite{szegedy2016inception} which aims to take into consideration the context of the whole image input. The decoder is composed of the series of the transpose convolution layers that have PMM modules and dual-attentions incorporated within the decoder blocks. The layer architecture of the proposed PMAD-LinkNet is presented in Figure \ref{fig:segArch}.

\subsubsection{PMAD-LinkNet Encoder}
The encoder section of the segmentation architecture consists of a stem block, three types of InceptionResNet blocks, and two types of reduction blocks.

The stem block starts off with a series of three convolutions before proceeding to a max pooling layer and a convolution layer which occur in parallel. After that comes a filter concatenation layer which in turn branches out into two parallel paths. One of the branches has two convolutional layers while the other branch has four of them. Both branches rejoin through filter concatenation and are in turn followed again by a parallel convolution and max pooling and then another filter concatenation step.

\begin{equation} \label{eq:conv}
\underbrace{\text{Conv}(I_{(i,j)}, F)}_{\text{Convolutional Operation}} = \sum_{m=0}^{M-1} \sum_{n=0}^{N-1} \underbrace{I_{(i+m,j+n)}}_{\text{Input feature map}} \cdot \underbrace{F_{(m,n)}}_{\text{Filter}} + \underbrace{b}_{\text{Bias}}
\end{equation}

The Inception Resnet blocks are of three types, named A, B and C respectively. Block A is composed of three different paths and a residual connection. The first path consists of a single convolution operation while the second and third paths consist of three and two convolutional operations respectively. The three paths are combined with the help of another convolution operation followed by concatenation with the residual connection. Blocks B and C are similar but the major difference is with the size of the feature maps since an average pooling operation is responsible for downsampling the data from block B to block C. They are composed of two different paths, one with three convolution operations and the other with one convolution operation. The convolution paths are combined by utilizing another convolution operation. There also exists a residual connection which is combined with the result of the convolution operations by utilizing a convolution operation.

\begin{equation}
\underbrace{\text{Concatenation}(A, B)_{(i,j,k)}}_{\text{Concatenation Operation}} =
\begin{cases}
\underbrace{A_{(i,j,k)}}_{\text{Input feature map}} & \text{if } 1 \leq k \leq \text{depth}(A) \\
\underbrace{B_{(i,j,k-\text{depth}(A))}}_{\text{Input feature map}} & \text{if } \text{depth}(A) \leq k \leq \text{depth}(A) + \text{depth}(B)
\end{cases}
\end{equation}

\begin{equation}
\underbrace{\text{ReLU}(x)}_{\text{Rectified Linear Unit Activation}} =
\begin{cases}
\underbrace{x}_{\text{Feature map}} & \text{if } x > 0 \\
0 & \text{otherwise}
\end{cases}
\end{equation}

The Reduction blocks are of two variants which are called A and B respectively. Block A begins with a filter concatenation operation which is then split into three paths. The first and third paths are composed of a max pooling and a convolution operation respectively and the second path is composed of three convolution layers. The three paths are then combined with the help of a filter concatenation operation. Block B also consists of a max pooling operation and three convolution operations which are present parallelly. Unlike block A, block B is composed of four different parallel paths where the first two paths are described in the previous statement. The other two paths are two convolution operations respectively and all the four paths are combined by utilising a filter concatenation operation.

\begin{equation}
\underbrace{\text{MaxPooling}(O)_{(i,j)}}_{\text{Max Pooling Operation}} = \max_{p=0}^{k-1} \max_{q=0}^{k-1} \underbrace{I_{(i \cdot s + p, j \cdot s + q)}}_{\text{Input feature map}}
\end{equation}

\begin{equation}
\underbrace{\text{FilterConcat}(F_1, F_2)(X)}_{\text{Filter Concatenation}} = \underbrace{\text{Concatenation}(\text{Conv}(X, F_1), \text{Conv}(X, F_2))}_{\text{Concatenated Convolutions}}
\end{equation}


\subsubsection{PMAD-LinkNet Decoder}

The decoder section of the PMAD-LinkNet segmentation framework integrates the PMM module within its decoder blocks to achieve precise pixel-level refinement of tumor boundaries. Each decoder block comprises convolutional layers, batch normalization, transpose convolution layers, spatial-channel attention blocks, and the PMM module, culminating with a softmax activation function for the final segmentation output.

The decoder block begins with a convolution operation followed by batch normalization, as described in Equation~\eqref{eq:batch_norm}. Batch normalization stabilizes the learning process by normalizing the input features using the mean and variance computed over the mini-batch:

\begin{equation}
\underbrace{\text{BN}(x)}_{\text{Batch Normalization}} = \underbrace{\gamma}_{\text{Learnable parameter}} \left( \frac{\underbrace{x}_{\text{Input}} - \underbrace{\mu}_{\text{Mean}}}{\sqrt{\underbrace{\sigma^2}_{\text{Variance}} + \underbrace{\epsilon}_{\text{Small constant}}}} \right) + \underbrace{\beta}_{\text{Learnable parameter}}
\label{eq:batch_norm}
\end{equation}

A transpose convolution operation follows, which upsamples the feature maps to recover spatial resolution lost during encoding. This operation is mathematically represented in Equation~\eqref{eq:transpose_conv}:

\begin{equation}
\underbrace{\text{TransposeConv}(X, K)_{(i,j,d)}}_{\text{Transpose Convolution}} = \sum_{p=0}^{F-1} \sum_{q=0}^{F-1} \sum_{c=0}^{C-1} \underbrace{X_{(i+s \cdot p, j+s \cdot q, c)}}_{\text{Input feature map}} \cdot \underbrace{K_{(p, q, c, d)}}_{\text{Filter kernel}}
\label{eq:transpose_conv}
\end{equation}

After another batch normalization, the \textbf{Precision Mapping Mechanism (PMM)} is introduced. The PMM dynamically calibrates spatial mappings through a spatial-channel attention block by analyzing morphological variations within the feature maps. It computes adaptive spatial adjustments for each pixel, effectively refining the feature map to accurately localize tumor edges. This is achieved through a learned refinement process that compensates for inconsistencies in image quality and variations in tumor morphology. It ensures that the spatial mappings are precisely aligned with the tumor boundaries at the pixel level. It begins with two parallel double convolution operations, capturing diverse spatial features. The resulting feature maps are then combined using an addition operation as per Equation~\eqref{eq:addition}. A ReLU activation introduces non-linearity, followed by another convolution operation. The output is passed through a sigmoid activation function (Equation~\eqref{eq:sigmoid}) to normalize the feature values between 0 and 1.

\begin{equation}
\underbrace{\text{Addition}(A, B)_{(i,j)}}_{\text{Addition Operation}} = \underbrace{A_{(i,j)}}_{\text{Input feature map}} + \underbrace{B_{(i,j)}}_{\text{Input feature map}}
\label{eq:addition}
\end{equation}

\begin{equation}
\underbrace{\text{Sigmoid}(x)}_{\text{Sigmoid Activation}} = \frac{1}{1 + e^{- \underbrace{x}_{\text{Input}}}}
\label{eq:sigmoid}
\end{equation}

\begin{equation}
\underbrace{\text{AveragePooling}(X)_{(i,j)}}_{\text{Average Pooling}} = \frac{1}{k^2} \sum_{p=0}^{k-1} \sum_{q=0}^{k-1} \underbrace{X_{(i \cdot s + p, j \cdot s + q)}}_{\text{Input feature map}}
\label{eq:average_pooling}
\end{equation}

The pooled features are input to a shared multi-layer perceptron (MLP) consisting of flatten layers, Gaussian Error Linear Unit (GELU) activation functions, and dropout layers. The GELU activation function is defined in Equations~\eqref{eq:gelu1} and~\eqref{eq:gelu2}. Dropout regularization is applied to prevent overfitting, as shown in Equation~\eqref{eq:dropout}.

\begin{equation}
y = \underbrace{\text{GELU}(W \cdot x + b)}_{\text{Gated Linear Unit}}
\label{eq:gelu1}
\end{equation}

\begin{equation}
\underbrace{\text{GELU}(x)}_{\text{Gaussian Error Linear Unit}} = x \cdot \underbrace{\Phi(x)}_{\text{Gaussian error function}}
\label{eq:gelu2}
\end{equation}

\begin{equation}
\underbrace{O_{(i,j)}}_{\text{Output feature map}} = \frac{\underbrace{I_{(i,j)}}_{\text{Input feature map}}}{1 - \underbrace{\text{rate}}_{\text{Dropout rate}}}
\label{eq:dropout}
\end{equation}

The PMM reweights the feature maps, emphasizing the most informative features for segmentation. The decoder operation concludes with a transpose convolution, two convolution operations, and a softmax activation function ~\eqref{eq:softmax} to generate the segmented output.

\begin{equation}
\underbrace{\text{softmax}(z)}_{\text{Softmax Activation}} = \frac{e^{\underbrace{z}_{\text{Output score}}}}{\sum_{i,j} e^{z}}
\label{eq:softmax}
\end{equation}

By integrating the PMM within the decoder blocks, the PMAD-LinkNet framework dynamically adapts spatial mappings based on morphological variations. The PMM's adaptive mapping strategy provides pixel-level precision in localizing tumor edges by adjusting spatial mappings through learned refinements. This approach effectively compensates for inconsistencies in image quality and variations in tumor morphology, leading to enhanced segmentation accuracy and precise delineation of tumor boundaries. The segmentation process is algorithmically presented in Algorithm \ref{alg:segmentation}. 

\begin{algorithm}
\caption{PMAD-LinkNet for Breast Lesion Segmentation}
\label{alg:segmentation}
\begin{algorithmic}[1]
\REQUIRE Preprocessed breast ultrasound images $\text{breastUltrasound}_{\text{preprocessed}}$, ground truth masks $\text{ground\_truth}$
\ENSURE Trained segmentation framework $\text{model}_{\text{segmentation}}$
\STATE
\STATE \textbf{Function} \texttt{trainFramework}($\text{breastUltrasound}_{\text{preprocessed}}$, $\text{ground\_truth}$):
\STATE $\beta \leftarrow$ batch size
\STATE $T \leftarrow$ total epochs
\STATE $\theta \leftarrow$ parameters of $\text{model}_{\text{segmentation}}$
\STATE $n \leftarrow$ total classes
\STATE Define loss functions:
\STATE \quad $\mathcal{L}_{1} \leftarrow$ Jaccard loss
\STATE \quad $\mathcal{L}_{2} \leftarrow$ Dice loss
\STATE \quad $\mathcal{L}_{3} \leftarrow$ Categorical Focal loss
\FOR{epoch $\leftarrow 1$ \TO $T$}
    \STATE $\mu \leftarrow$ learning rate
    \FOR{each batch $(x, \text{mask}_{\text{GT}})$ sampled from training data}
        \STATE $\hat{y} \leftarrow \text{model}_{\text{segmentation}}(x)$
        \STATE $\mathcal{L}_{\text{total}} \leftarrow \mathcal{L}_{1}(\hat{y}, \text{mask}_{\text{GT}}) + \mathcal{L}_{2}(\hat{y}, \text{mask}_{\text{GT}}) + \mathcal{L}_{3}(\hat{y}, \text{mask}_{\text{GT}})$
        \STATE Update $\theta$ using backpropagation with loss $\mathcal{L}_{\text{total}}$:
        \STATE $\theta \leftarrow \theta - \mu \nabla_{\theta} \mathcal{L}_{\text{total}}$
        \COMMENT{where $\nabla_{\theta} \mathcal{L}_{\text{total}}$ is the gradient of the loss with respect to $\theta$}
    \ENDFOR
    \IF{validation performance plateaus}
        \STATE Adjust learning rate $\mu$ to promote further learning
    \ENDIF
\ENDFOR
\STATE $\text{output}_{\text{segmentation}} \leftarrow \text{model}_{\text{segmentation}}$
\RETURN $\text{output}_{\text{segmentation}}$
\end{algorithmic}
\end{algorithm}

\subsubsection{Training Setup and Specifications}

The model was trained by backpropagating over a custom loss function (Equation \ref{eq:segLoss}), equal to an aggregate of focal loss (Equation \ref{eq:focLoss}) and dice (Jaccard) loss (Equation \ref{eq:jacLoss}) obtained after each training epoch.

\begin{equation}\label{eq:focLoss}
\underbrace{\text{loss}_\text{focal}(p_t)}_{\text{Focal loss}} = - (1 - \underbrace{p_t}_{\text{True class probability}})^{\underbrace{\gamma}_{\text{Focal loss focusing parameter}}} \log(\underbrace{p_t}_{\text{True class probability}})
\end{equation}

\begin{equation}\label{eq:jacLoss}
\underbrace{\text{loss}_\text{Jaccard}}_{\text{Jaccard loss}} = 1 - \frac{\underbrace{V_p \cap V_g}_{\text{Intersection of predicted and ground truth}}}{\underbrace{V_p \cup V_g}_{\text{Union of predicted and ground truth}}}
\end{equation}

\begin{equation} \label{eq:segLoss}
\underbrace{\text{loss}_\text{total}}_{\text{Total loss}} = \underbrace{\text{loss}_\text{focal}}_{\text{Focal loss}} + \underbrace{\text{loss}_\text{Jaccard}}_{\text{Jaccard loss}}
\end{equation}
The model specifications and parameters of the proposed PMAD-LinkNet segmentation framework are shown in Table \ref{tab:segSpec}.
 
\begin{table}[h!]
 \caption{PMAD-LinkNet Segmentation Framework Specifications}
  \centering
  \label{tab:segSpec}
  \begin{tabular}{ll}              \\
    \cmidrule(r){1-2}
    \textbf{Parameters}     & \textbf{Coefficients}   \\
    \midrule
    Total Trainable Parameters & 57,881,011            \\
Learning Rate              & 0.001                 \\
Epochs                     & 100                   \\
Image Shape                & (256, 256)            \\
Batch Size                 & 16     \\
    \bottomrule
  \end{tabular}
  \label{tab:table}
\end{table}
 
\subsection{Component-Specific Feature-Enhanced Classifier (CSFEC-Net) for Breast Cancer Classification}

This section presents the proposed Component-Specific Feature-Enhanced Classifier (CSFEC-Net) for breast lesion classification. The input to the model comprises segmented tumors and outputs the predicted class to which the lesion belongs.

\begin{figure} 
  \centering

  \includegraphics[width=0.75\textwidth]{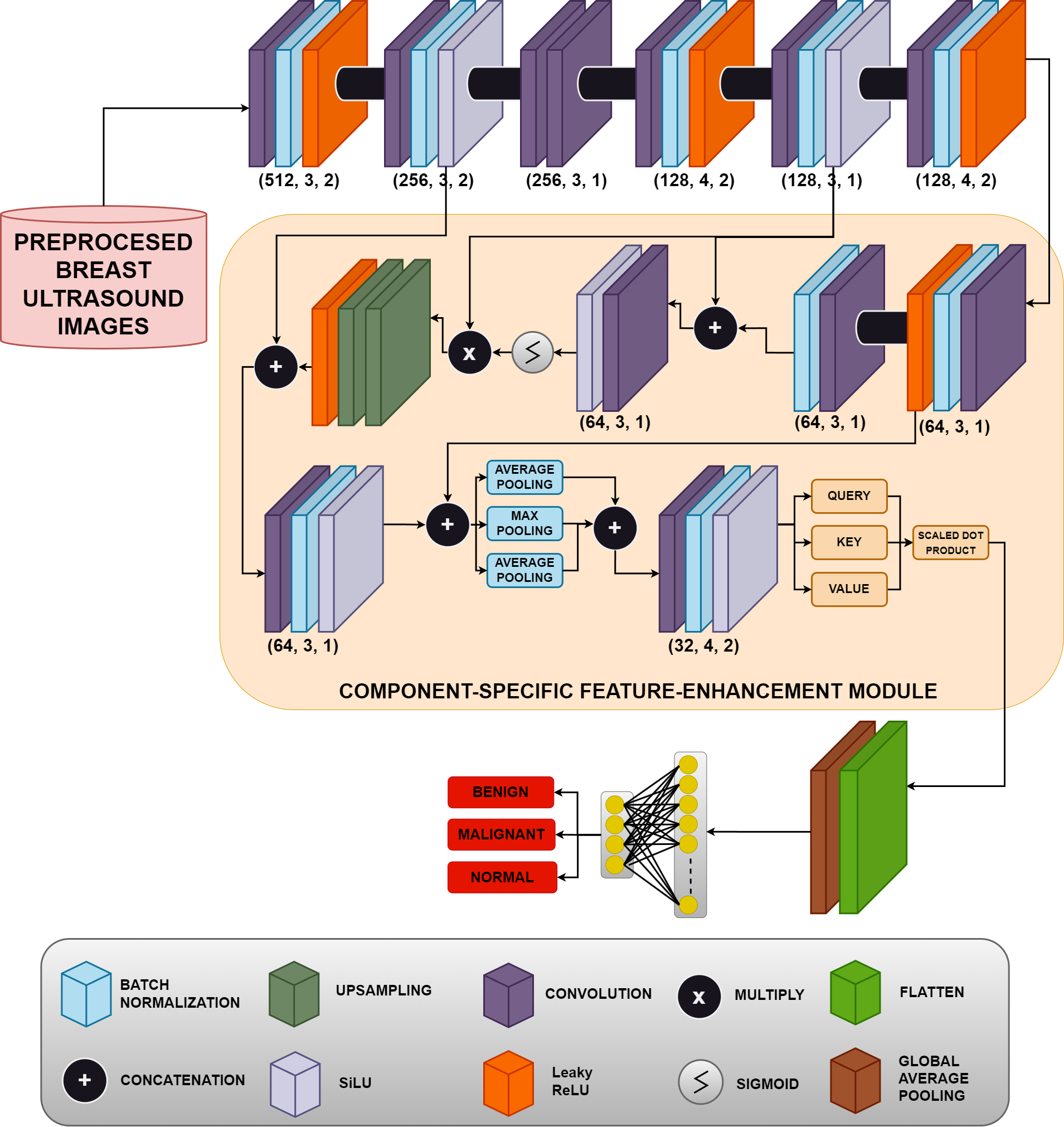}
  \caption{CSFEC-Net Classification Layer Architecture for Breast Cancer Classification}

    \label{fig:clsArch}
\end{figure}

The model architecture of CSFEC-Net integrates several pivotal blocks designed to extract pertinent features from the input breast ultrasound images. These blocks include convolutional blocks, double convolutional blocks, self-attention blocks, and fully connected layers. Each block plays a crucial role in feature extraction and classification. The layer architecture diagram of the proposed CSFEC-Net model is shown in Figure \ref{fig:clsArch}.

\subsubsection{Workflow and Execution}

The classification module of the Component-Specific Feature-Enhanced Classifier (CSFEC-Net) is meticulously designed to accurately differentiate between malignant, benign, and normal breast tissues. Central to this module is the novel Component-Specific Feature Enhancement Module (CSFEM), which employs dynamic multi-attention mechanisms to selectively amplify discriminative features across different tissue types. The information flow within the CSFEC-Net commences with an input layer of dimensions \(256 \times 256 \times 3\), corresponding to the input segmented tumor. 

Initially, the segmentation map is processed through a convolutional block (\ref{eq:conv}) comprising 512 filters, a padding of 2, and a kernel size of 3. This block extracts low-level features such as textures and edges from the input image. The output from the first convolutional block is then passed through a second convolutional block with 256 filters, identical kernel size, and padding, but incorporates the Sigmoid Linear Unit (SiLU) activation function (\ref{eq:SiLU}) to introduce non-linearity for enhanced feature discrimination.

\begin{equation} \label{eq:SiLU}
\underbrace{\text{SiLU}(x)}_{\text{Sigmoid Linear Unit}} = \frac{x}{1 + e^{-x}}
\end{equation}

A double convolutional block is then employed, consisting of two consecutive convolutional layers with 256 filters each. This sequential processing facilitates the extraction of deeper and more abstract features, which are crucial for distinguishing between different breast tissue characteristics indicative of cancerous growth. The integration of the Component-Specific Feature Enhancement Module (CSFEM) follows, dynamically enhancing discriminative features specific to benign, malignant, and normal tissues. The CSFEM employs a multi-level attention approach, leveraging both spatial and self attention mechanisms to magnify distinguishing features. The attention-enhanced feature map \( F' \) is computed as:
\begin{equation}
F' = F \otimes \text{SA}(F) \otimes \text{CA}(F)
\label{eq:attention}
\end{equation}
where \( F \) represents the input feature map, \( \text{SA}(F) \) denotes the spatial attention map, \( \text{CA}(F) \) denotes the channel attention map and \( \otimes \) signifies element-wise multiplication. To capture features at multiple scales, the CSFEM incorporates a multi-scale attention pipeline:
\begin{equation}
\text{MSA}(F) = \text{Concat}(\text{Conv}(F, 1), \text{Conv}(F, 3), \text{Conv}(F, 5)) \cdot W
\label{eq:msa}
\end{equation}
where \( \text{Conv}(F, k) \) denotes a convolution operation with kernel size \( k \), and \( W \) represents the learnable weights for feature aggregation. The enhanced features are obtained by applying the multi-scale attention maps to the original feature maps:
\begin{equation}
F_{\text{enhanced}} = F + \text{MSA}(F) \otimes F
\label{eq:feature_enhance}
\end{equation}
This process increases inter-class separability and intra-class consistency by selectively amplifying discriminative traits of different tissue types, effectively addressing overlapping feature spaces in heterogeneous tumor segmentations. It enhances the model's ability to focus on relevant cellular distributions and the presence of necrosis, filtering out irrelevant information and potential artifacts that could obscure diagnosis. 

Following the spatial attention in the CSFEM, the feature maps undergo additional convolutional blocks to further distill discriminative representations. A convolutional block with 128 filters, a kernel size of 4, and padding of 2, coupled with a Leaky ReLU activation function (\ref{eq:leakyrelu}), refine the features. Subsequently, another convolutional block with 128 filters, padding of 1, and a kernel size of 3 using SiLU activation is applied. Two additional convolutional blocks with 128 and 64 filters, kernel sizes of 4 and 3, and appropriate padding are utilized before feeding the features into a spatial attention mechanism. This mechanism adjusts the feature maps to accentuate subtle differences between various tissue characteristics associated with malignant and benign tumors. 

\begin{equation}\label{eq:leakyrelu}
\text{LeakyReLU}(x) = 
\begin{cases} 
\underbrace{x}_{\text{Input value}} & \text{if } x > 0 \\ 
\underbrace{\alpha x}_{\text{Leaky slope}} & \text{otherwise}
\end{cases}
\end{equation}

The feature map from the spatial attention mechanism of the CSFEM is concatenated with the output from a convolution and batch normalization (\ref{eq:batch_norm}) layer with 64 filters, padding of 2, and a kernel size of 3. This concatenation integrates both high-level and low-level features across different layers, enabling a more comprehensive representation of the input image. The enriched feature map is further processed through convolutional and activation layers before being upsampled and concatenated with intricate feature attention results. This iterative refinement ensures that the model effectively leverages both global and local contextual details present in the input segmentation map.

Following this, the feature map is flattened and subjected to dropout regularization to mitigate overfitting. Dropout prevents the model from relying on specific features or patterns within the training data that may not generalize well to unseen samples, thereby improving its robustness and generalization performance. The dropout-regularized features are then directed into a fully connected layer containing 128 neurons, which transforms the features into a compact representation suitable for classification. Finally, the output layer consists of three neurons with a softmax activation function (\ref{eq:softmax}) to classify the input into malignant, benign, or normal categories. The algorithmic workflow of our CSFEC-Net is provided in Algorithm \ref{alg:csfam-net}. 

\begin{algorithm}
\caption{CSFEC-Net for Breast Cancer Classification}
\label{alg:csfam-net}
\begin{algorithmic}[1]
\REQUIRE Segmented breast tumor images $\text{breastTumor}_{\text{segmented}}$
\ENSURE Trained CSFEC-Net classifier model $\text{model}_{\text{classification}}$
\STATE
\STATE \textbf{Function} \texttt{TrainClassifier}($\text{breastTumor}_{\text{segmented}}$):
\STATE Initialize parameters $\theta$ of $\text{model}_{\text{classification}}$
\STATE $\beta \leftarrow$ batch size
\STATE $T \leftarrow$ total epochs
\FOR{epoch $\leftarrow 1$ \TO $T$}
    \STATE $\mu \leftarrow$ learning rate
    \FOR{each batch $(x, y)$ sampled from training data}
        \STATE $\hat{y} \leftarrow \text{model}_{\text{classification}}(x)$
        \STATE $\mathcal{L}_{\text{loss}} \leftarrow$ categorical cross-entropy loss between $y$ and $\hat{y}$
        \STATE Update $\theta$ using backpropagation with loss $\mathcal{L}_{\text{loss}}$:
        \STATE $\theta \leftarrow \theta - \mu \nabla_{\theta} \mathcal{L}_{\text{loss}}$
        \COMMENT{where $\nabla_{\theta} \mathcal{L}_{\text{loss}}$ is the gradient of the loss with respect to $\theta$}
    \ENDFOR
    \IF{validation performance plateaus}
        \STATE Adjust learning rate $\mu$ to promote further learning
    \ENDIF
\ENDFOR
\STATE $\text{output}_{\text{DCNNIMAF}} \leftarrow \text{model}_{\text{classification}}$
\RETURN $\text{output}_{\text{DCNNIMAF}}$
\end{algorithmic}
\end{algorithm}

\subsubsection{Training Setup and Specifications}
Our proposed CSFEC-Net’s training parameters were updated after each epoch via backpropagation using the categorical cross entropy loss criterion (\ref{eq:CCE}). 

\begin{equation} \label{eq:CCE}
\underbrace{\text{loss}_\text{CE}}_{\text{Categorical Cross Entropy Loss}} = - \log \left( \frac{\underbrace{e^{x_p}}_{\text{Exponential of the true class score}}}{\underbrace{\sum_{j=1}^N e^{x_j}}_{\text{Sum of exponentials of all class scores}}} \right)
\end{equation}

The model specifications and parameters of the proposed CSFEC-Net classifier have been shown in Table \ref{tab:clsSpec}.

\begin{table}[h!]
 \caption{Specifications of the Proposed CSFEC-Net Classification Model}
  \centering
  \begin{tabular}{ll}              \\
    \cmidrule(r){1-2}
    \textbf{Parameters}     & \textbf{Coefficients}   \\
    \midrule
    Total Trainable Parameters & 52,427,081            \\
Learning Rate              & 0.001                 \\
Epochs                     & 100                   \\
Image Shape                & (256, 256)            \\
Batch Size                 & 16     \\
    \bottomrule
  \end{tabular}
  \label{tab:clsSpec}
\end{table}

\section{Experimental Setup and Results}

This section presents the results and arguments derived from the application of the proposed models. The experiment was realized on a machine with the following infrastructure: CPU - AMD Ryzen 7 4800H with Radeon Graphics, x86\_64 architecture, at a speed of 3GHz with 8 cores; GPU - NVIDIA GeForce RTX 3050-PCI Bus 1; and RAM is 32 GB. These points are summarized in Table \ref{tab:trainingSpecs}. 

\begin{table}[h!]
 \caption{System Specifications for Experimental Setup}
  \centering
  \begin{tabular}{ll}              \\
    \cmidrule(r){1-2}
    \textbf{Component}     & \textbf{Specification}   \\
    \midrule
    CPU          & AMD Ryzen 7 4800H with Radeon Graphics \\
    ARCHITECTURE & x86\_64                                \\
    BASE SPEED   & 3GHz                                   \\
    CORES        & 8                                      \\
    GPU          & NVIDIA GeForce RTX 3050-PCI Bus 1      \\
    RAM          & 32GB                                   \\
    \bottomrule
  \end{tabular}
  \label{tab:trainingSpecs}
\end{table}

\subsection{Segmentation Evaluation Metrics}

The evaluations in terms of the proposed segmentation framework were carried out in the training and validation phase using the following metrics entailing segmentation:

\subsubsection{Accuracy}
Accuracy measures the proportion of pixels that were classified correctly in the segmentation map compared to the ground truth.

\begin{equation}
\underbrace{\text{accuracy}}_{\text{Segmentation Accuracy}} = \frac{\underbrace{\text{correctly\_classified\_pixels}}_{\text{Number of correctly classified pixels}}}{\underbrace{\text{total\_pixel\_count}}_{\text{Total number of pixels in the image}}}
\end{equation}

\subsubsection{IoU Score}
The IoU score, often termed the Jaccard index, assesses the intersection of the ground truth mask with the predicted segmentation mask divided by their union. It represents the amount of tumor region correctly segmented regarding the total tumor region (ground truth).

\begin{equation}
\underbrace{\text{IoU}_{\text{Score}}}_{\text{IoU}} = \frac{\underbrace{\text{Area}_{\text{segmentation}} \cap \text{Area}_{\text{groundTruth}}}_{\text{Intersection}}}{\underbrace{\text{Area}_{\text{segmentation}} \cup \text{Area}_{\text{groundTruth}}}_{\text{Union}}}
\end{equation}

\subsubsection{Dice Coefficient}
The Dice coefficient, often recognized as the Dice similarity index, assesses the overlap between the ground truth and the predicted segmentation mask.

\begin{equation}
\underbrace{\text{Dice}_{\text{Coefficient}}}_{\text{Dice}} = \frac{2 \times |\underbrace{\text{Area}_{\text{segmentation}} \cap \text{Area}_{\text{groundTruth}}}_{\text{Intersection}}|}{|\underbrace{\text{Area}_{\text{segmentation}}}_{\text{Segmentation}}| + |\underbrace{\text{Area}_{\text{groundTruth}}}_{\text{Ground Truth}}|}
\end{equation}

\begin{figure}[htbp]
    \centering
    \begin{minipage}[b]{0.48\textwidth}
        \centering
        \includegraphics[width=0.8\textwidth]{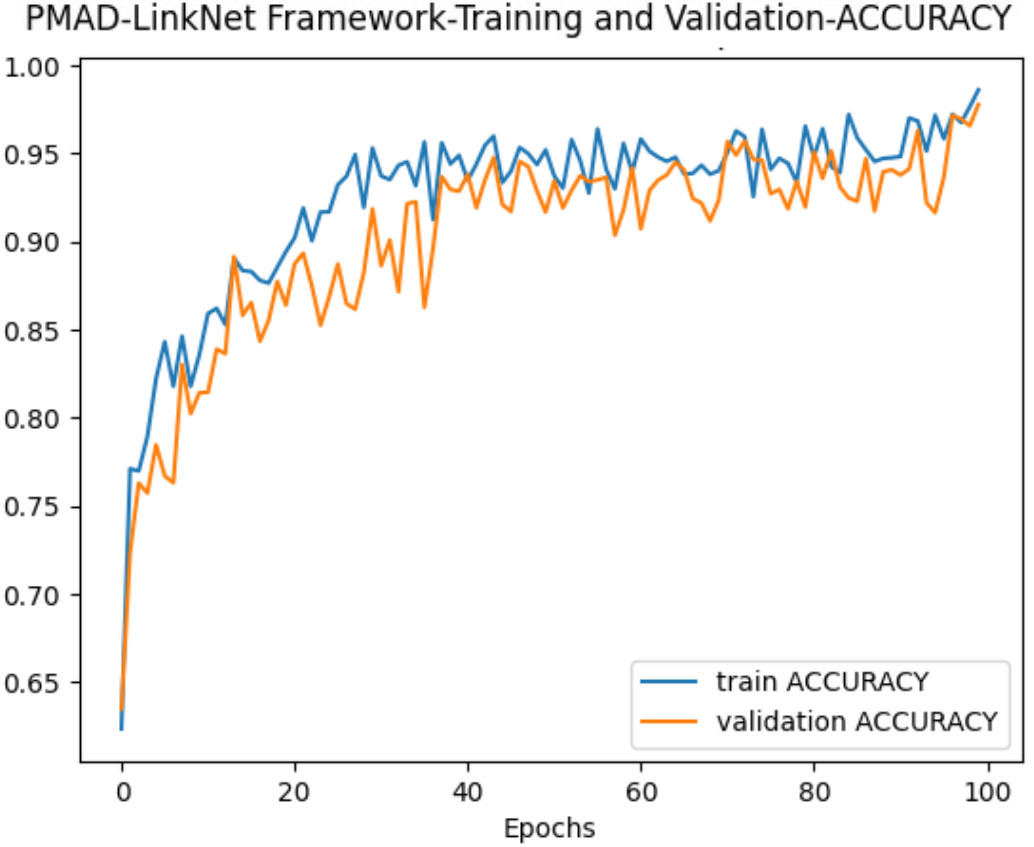} 
        \caption{Training and Validation Accuracy Curves of Proposed Segmentation Framework}
        \label{fig:img1}
    \end{minipage}
    \hfill
    \begin{minipage}[b]{0.48\textwidth}
        \centering
        \includegraphics[width=0.8\textwidth]{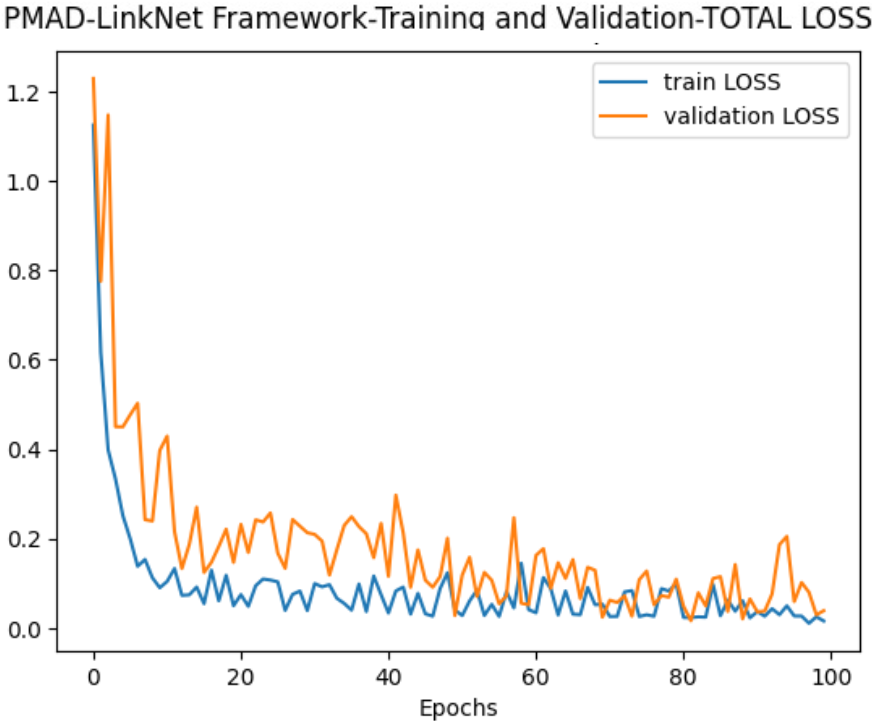} 
        \caption{Training and Validation Loss Curves of Proposed Segmentation Framework}
        \label{fig:img2}
    \end{minipage}
\end{figure}

\begin{figure}[htbp]
    \centering
    \begin{minipage}[b]{0.48\textwidth}
        \centering
        \includegraphics[width=0.8\textwidth]{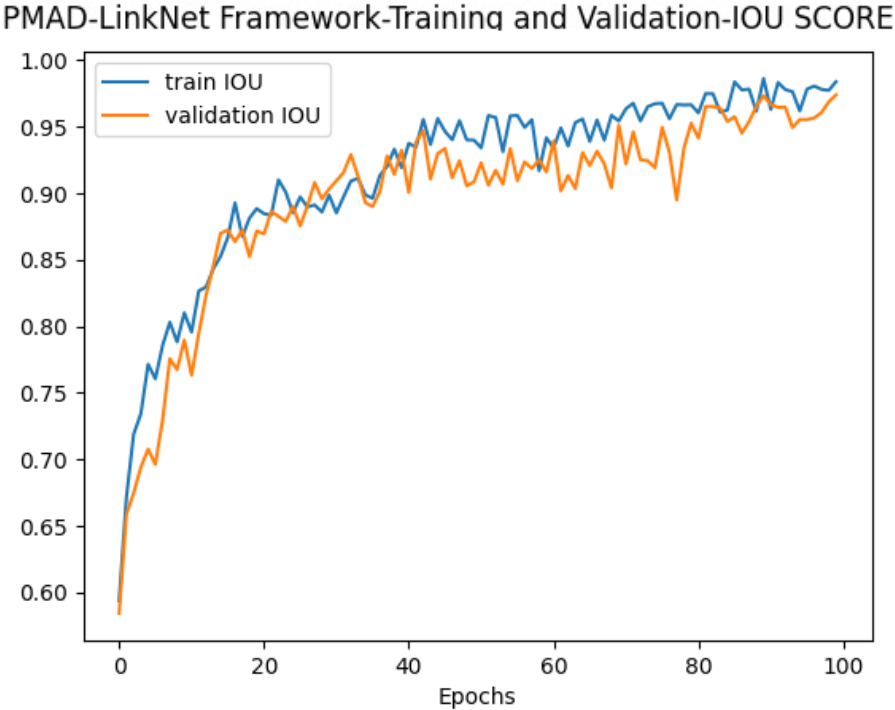} 
        \caption{Training and Validation IoU Score Curves of Proposed Segmentation Framework}
        \label{fig:img3}
    \end{minipage}
    \hfill
    \begin{minipage}[b]{0.48\textwidth}
        \centering
        \includegraphics[width=0.8\textwidth]{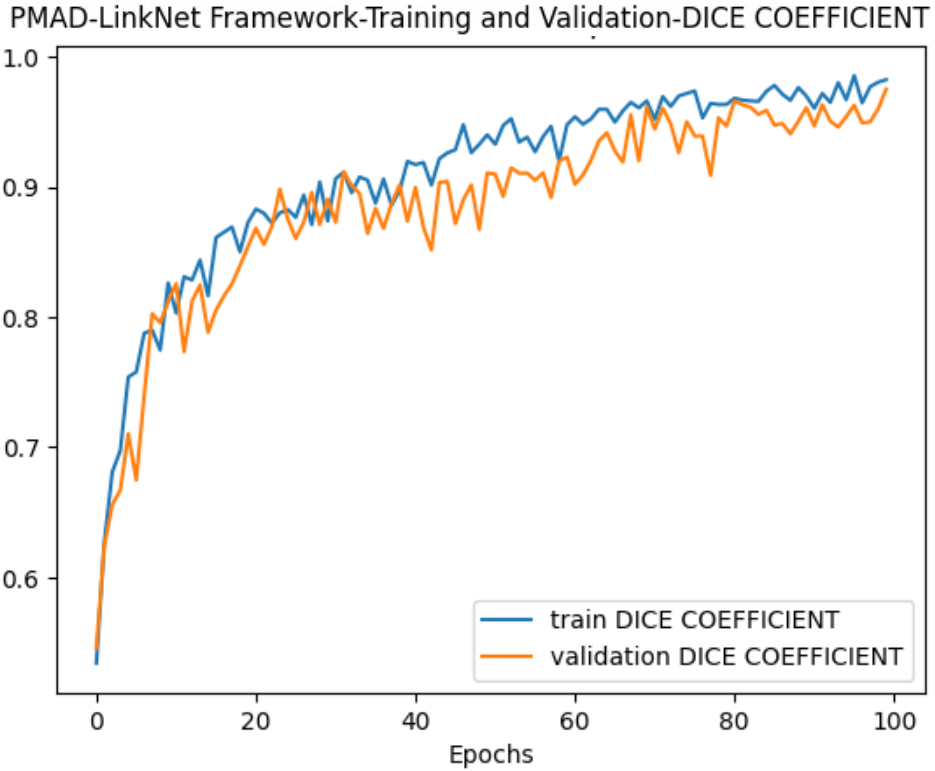} 
        \caption{Training and Validation Dice Coefficient Curves of Proposed Segmentation Framework}
        \label{fig:img4}
    \end{minipage}
\end{figure}

Figures \ref{fig:img1} and \ref{fig:img2} depict the training and validation curves for accuracy and total loss, respectively, obtained while training the proposed segmentation framework. From the graphs, it is evident that the model has achieved a high accuracy of 98.1\%, with a minimal loss of 0.06 at the end of 100 epochs. The model also achieved an impressive Dice Coefficient score of 97.2\% and an IoU score of 96.9\%. The training and validation curves of these metrics have been shown in Figures \ref{fig:img3} and \ref{fig:img4} respectively.

\subsubsection{Performance Evaluation and Discussion}

From the segmentation results, it can be extrapolated that the said model has performed remarkably well. High values of IoU, Dice Coefficient, and accuracy scores counterbalanced by very little total losses reflect that the InceptionResNet backbone is very good at extracting valuable features out of input preprocessed images and the PMM module inserted in the decoder blocks fine-tunes the segmentation maps for segmentation \cite{selvaraju2017gradcam}.

Grad-CAM stands for gradient-weighted class activation mapping and is an algorithm denoting a methodology applied in DL for visualizing the different important areas that could influence a model's decision in an input image. The primary area of this research is how the CNNs make predictions in different tasks and also applies in medical image segmentation wherein the attention mechanism must be observed if it does its operations correctly.

\begin{figure}[h!]
  \centering
  \includegraphics[width=0.7\textwidth]{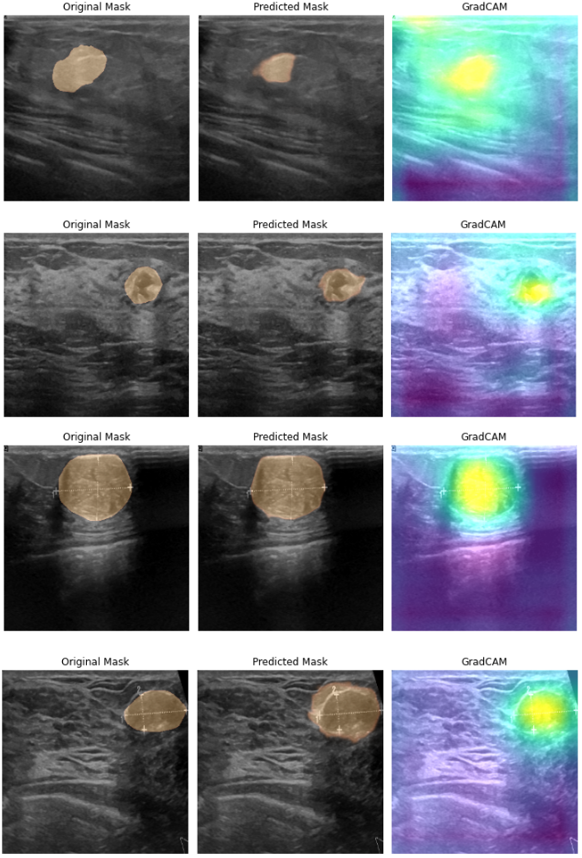}
  \caption{Segmentation Outputs with Attention GradCAMs at epochs 16, 32, 64, and 96}
  \label{fig:gc}
\end{figure}

The GradCAMs with respect to the attention block of the topmost decoder block as illustrated by Figure \ref{fig:gc} shows us how the attention mechanism can focus on certain regions within the feature map and highlight their importance towards the segmentation task. This visualization can illustrate that the attention mechanism contributes to the segmentation performance by highlighting the most relevant features with respect to their most important spatial locations. From GradCAMs, it can be observed that the attention mechanism progressively shifts its focus towards the tumor region, with an improvement in localization accuracy as the number of training epochs increases. Table \ref{tab:segComp} compares the performances of existing literature in breast lesion segmentation against our proposed PMAD-LinkNet.

\begin{table}[h!]
 \caption{Performance Metrics Comparison of Proposed Segmentation Model with Other Models}
  \centering
  \begin{tabular}{p{5cm}ll}
    \toprule
    & \multicolumn{2}{c}{\textbf{Performance Scores (in \%)}} \\
    \cmidrule(r){2-3}
    \textbf{Segmentation Model} & \textbf{Dice Coefficient (\%)} & \textbf{IoU Score (\%)} \\
    \midrule
    U-Net \cite{almajalid2018development}& 82.52  & 69.76     \\
    Res-U-Net \cite{yue2022deep}& 88.01& 80.21\\
    U-Net with DenseNet backbone \cite{zhang2023fully} & 89.86& 79.12\\
    Multi-scale Fusion U-Net \cite{li2021multi}& 95.35       & 91.12 \\
    \textbf{PMAD-LinkNet (Proposed)}   & \textbf{97.20}& \textbf{96.91}\\
    \bottomrule
  \end{tabular}
  \label{tab:segComp}
\end{table}

From Table \ref{tab:segComp}, we can see that U-Net \cite{almajalid2018development} attains a Dice coefficient of 82.52\% and an IoU score of 69.76\%. These scores reflect a foundational capability in segmenting tumors from breast ultrasound images and highlight the model's limitations in capturing the full extent of tumor boundaries and internal structures, particularly in the nuanced textures and densities often found in breast tissues. Res U-Net \cite{yue2022deep} enhances the original U-Net with a Dice coefficient of 88\% and an IoU score of 80\%, demonstrating enhanced performance through the incorporation of residual connections, but further refinements in its network architecture and feature extraction are necessary to achieve optimal segmentation accuracy, especially in dealing with the variable echo intensities and shadowing effects commonly encountered in breast ultrasound imaging. By integrating a DenseNet backbone, the U-Net with DenseNet Backbone \cite{zhang2023fully} reaches a Dice coefficient of 89.8\% and an IoU score of 79.1\%, showcasing the benefits of dense connectivity in improving segmentation outcomes. However, additional strategies may be required to fully leverage the complex patterns inherent in breast ultrasound images, such as the differentiation between cystic and solid components of tumors, which is critical for accurate diagnosis. The Multi-scale Fusion U-Net \cite{li2021multi} achieves a Dice coefficient of 95.35\% and an IoU score of 91.12\%, marking a significant improvement over earlier models. But it shows suboptimal performance when handling the heterogeneity of breast tissues and the dynamic nature of tumor growth observed in ultrasound sequences. The proposed Spatial-Channel Attention LinkNet Framework with InceptionResNet Backbone stands out with a Dice coefficient of 97.20\% and an IoU score of 96.91\%. Feature analysis is carried out using spatial-channel attention mechanisms together with a robust InceptionResNet backbone, enabling precise localization and delineation of tumors, including distinguishing type of breast lesions based on texture, shape, and boundary characteristics.

\subsection{Classification Evaluation Metrics}

The classification performance outcomes that have been proposed by the CSFEC-Net model for breast cancer classification are measured during training and validation using the following classification metrics:

\subsubsection{Accuracy}
Accuracy is the basic measure for judging the performance of the model as a whole concerning all classes. It gives a proportion of true classifications (true positive and true negative) against the total number of images classified, thereby giving a picture of the overall performance of the model while classifying instances.
\begin{equation}
\text{Accuracy} = \frac{\sum_{k=1}^{n} (TP_k + TN_k)}{\sum_{k=1}^{n} (TP_k + TN_k + FP_k + FN_k)}
\end{equation}
Where:
\begin{itemize}
    \item \( TP \) represents the number of true positives.
    \item \( TN \) represents the number of true negatives.
    \item \( FP \) represents the number of false positives.
    \item \( FN \) represents the number of false negatives.
    \item \( n \) represents the total number of classes.
\end{itemize}

\subsubsection{Precision}
Precision focuses on the proportion of true positive predictions among all positive predictions made by the classifier. It is particularly important in situations where false positives are costly, as it helps in minimizing the impact of false positives on the overall performance of the model.
\begin{equation}
\text{Precision} = \frac{\sum_{k=1}^{n} TP_k}{\sum_{k=1}^{n} (TP_k + FP_k)}
\end{equation}

\subsubsection{Recall}
Recall is a measure of how well the classifier can find all the relevant instances that belong in a specific class. It is important when missing a positive instance (false negative) is far worse than classifying a negative one as positive (false positive). Recall helps ensure that the model has not left any important cases out.
\begin{equation}
\text{Recall} = \frac{\sum_{k=1}^{n} TP_k}{\sum_{k=1}^{n} (TP_k + FN_k)}
\end{equation}

\subsubsection{F1-Score}
F1-Score combines precision and recall into a single measure, providing a balanced view of the model's performance. It is useful in scenarios where both false positives and false negatives are equally important, and a balance between these two metrics is desired.
\begin{equation}
\text{F1 Score} = \frac{2 \sum_{k=1}^{n} TP_k}{\sum_{k=1}^{n} (2TP_k + TN_k + FP_k)}
\end{equation}

\begin{figure}[h!]
    \centering
    \begin{minipage}[b]{0.48\textwidth}
        \centering
        \includegraphics[width=0.93\textwidth]{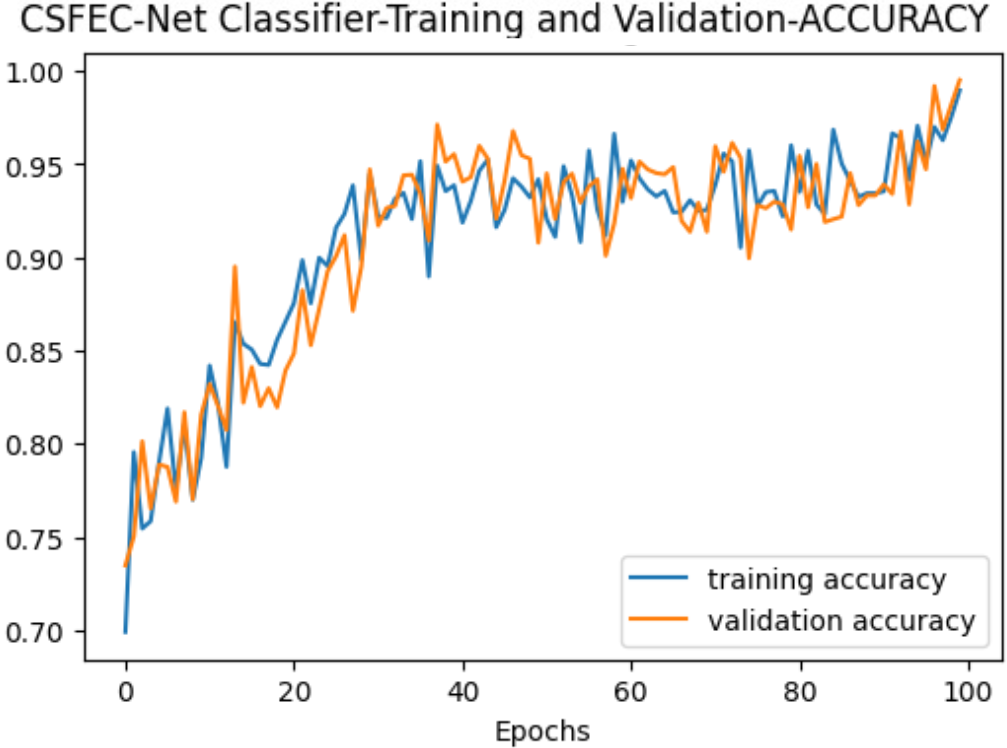} 
        \caption{Training and Validation Accuracy Curves of Proposed CSFEC-Net Classifier}
        \label{fig:img5}
    \end{minipage}
    \hfill
    \begin{minipage}[b]{0.48\textwidth}
        \centering
        \includegraphics[width=0.9\textwidth]{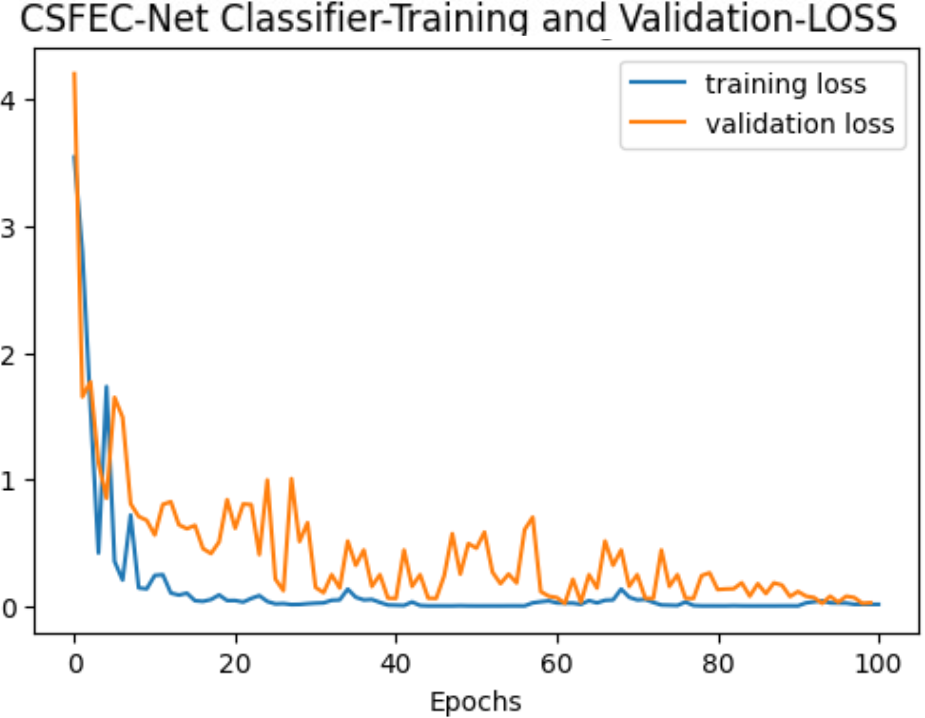} 
        \caption{Training and Validation Loss Curves of Proposed CSFEC-Net Classifier}
        \label{fig:img6}
    \end{minipage}
\end{figure}

\begin{figure}[h!]
    \centering
    \begin{minipage}[b]{0.48\textwidth}
        \centering
        \includegraphics[width=0.9\textwidth]{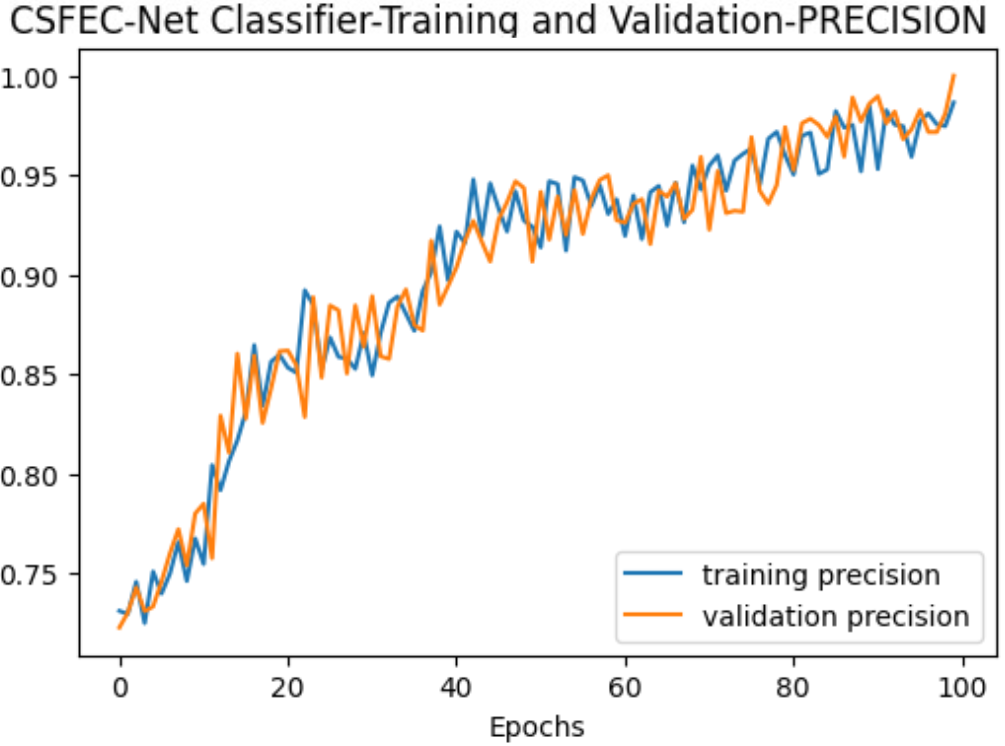} 
        \caption{Training and Validation Precision Curves of Proposed CSFEC-Net Classifier}
        \label{fig:img7}
    \end{minipage}
    \hfill
    \begin{minipage}[b]{0.48\textwidth}
        \centering
        \includegraphics[width=0.91\textwidth]{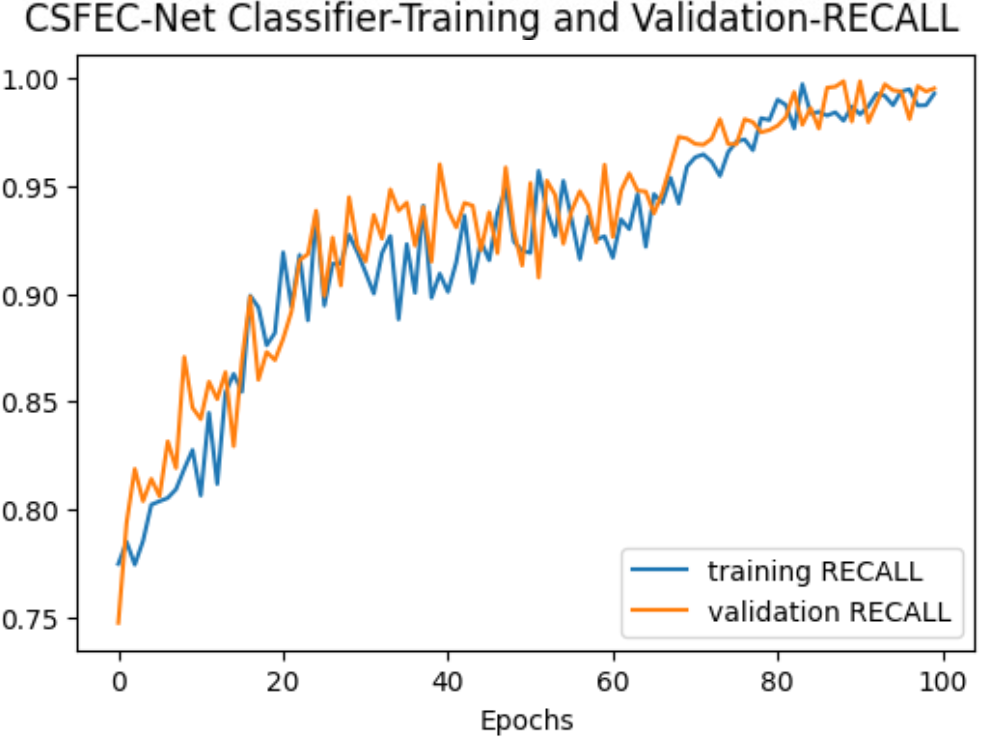} 
        \caption{Training and Validation Recall Curves of Proposed CSFEC-Net Classifier}
        \label{fig:img8}
    \end{minipage}
\end{figure}

\begin{figure}[h!]
  \centering
  \includegraphics[width=0.48\textwidth]{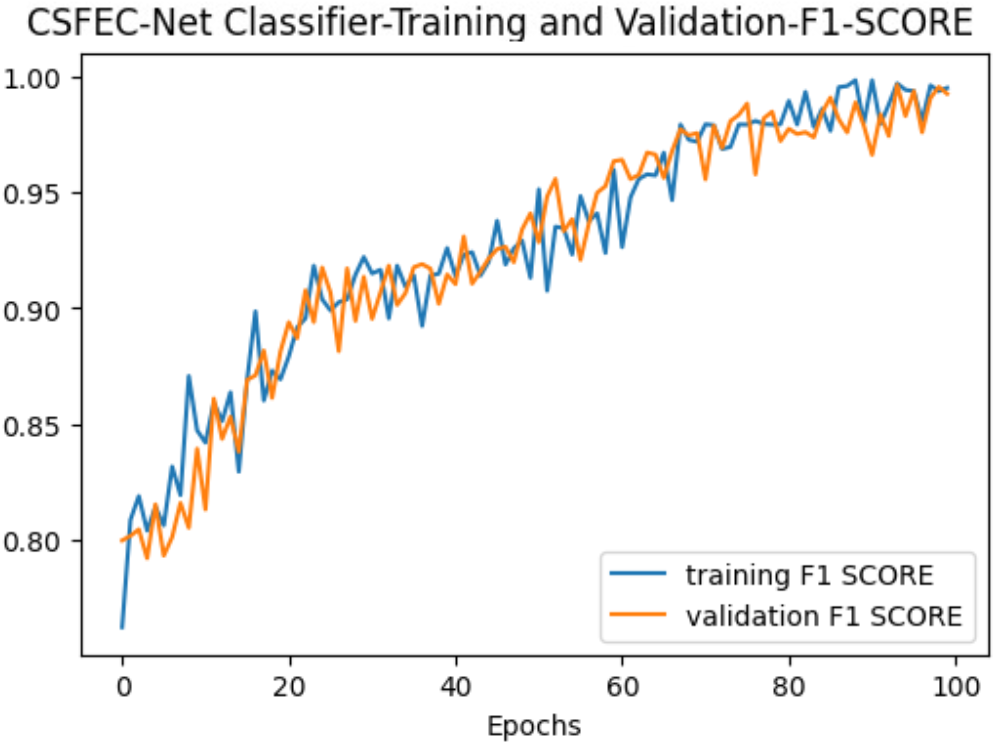}
  \caption{Training and Validation F1-Score Curves of Proposed CSFEC-Net Classifier}
  \label{fig:img9}
\end{figure}

The proposed CSFEC-Net classifier was trained for 100 epochs, and the evaluation metrics were recorded after each epoch. The training and validation curves obtained for accuracy coupled with categorical cross-entropy loss have been depicted in Figures \ref{fig:img5} and \ref{fig:img6} respectively. From the graph plots, it can be seen that the classification model has obtained a high accuracy of 99.2\% at a minimal loss of 0.03. Figures \ref{fig:img7}, \ref{fig:img8}, and \ref{fig:img9} display the training and validation precision, recall, and F1-score curve, respectively. It can be inferred from the graphs, that the proposed model has minimized false positives and false negatives, thereby achieving a remarkable precision of 99.3\% and a recall of 99.1\%. The high values of precision and recall contribute to the high F1-score value of 99.1\%.


\subsubsection{Performance Evaluation and Discussion}

\begin{figure}[h!]
  \centering
  \includegraphics[width=0.48\textwidth]{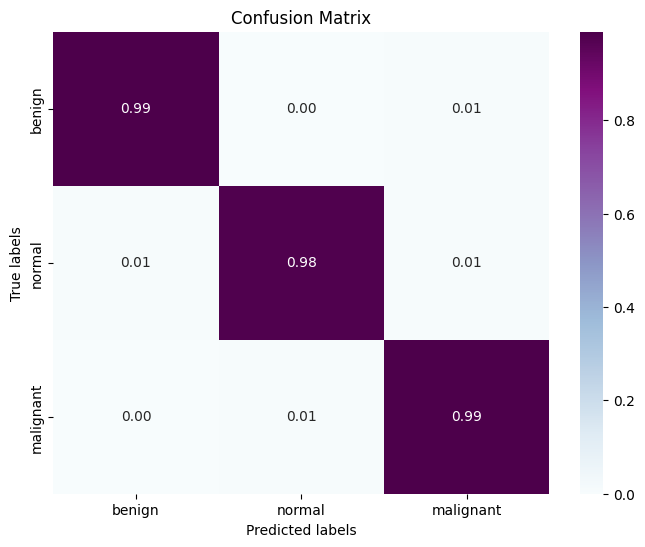}
  \caption{Confusion Matrix Obtained from Proposed CSFEC-Net Classifier}
  \label{fig:cm}
\end{figure}

The normalized confusion matrix obtained on the validation data using the trained CSFEC-Net classification model has been presented in Figure \ref{fig:cm}. A normalized confusion matrix is a type of confusion matrix where the values are normalized to show proportions or percentages. It is useful for comparing classification performance across classes, since the values are between 0 to 1, making it easy to interpret.

In Figure 18, the normalized confusion matrix depicts the proposed model's classification performance across the three breast cancer classes: "benign," "normal," and "malignant." Each row corresponds to the actual class, with each column representing the predicted class. The matrix's values show the proportion of true-class cases that were successfully classified (along the diagonal) or misclassified (off-diagonal).

From the matrix, it can be observed that the model has obtained remarkable accuracy. With most values along the diagonal close to one, it indicates that the majority of the samples were categorized correctly. For the "benign" class, the model had a true positive rate of 0.99, indicating that 99\% of benign tumors were properly categorized. In the "normal" class, the true positive rate was 0.98, implying that 98\% of normal cases were correctly identified. Similarly, in the "malignant" class, the true positive rate was 0.99, indicating that 99\% of malignant tumors were correctly identified. Misclassification errors were minor, with extremely low false positive and false negative rates. 

The proposed CSFEC-Net model is compared with other pretrained CNNs, including EfficientNetV2\cite{tan2021efficientnetv2}, MobileNetV2 \cite{sandler2018inverted}, \cite{huang2016densely}, NASNetMobile\cite{zoph2017learning}, Xception\cite{chollet2016xception}, InceptionV3\cite{szegedy2015rethinking}, InceptionResNetV2\cite{szegedy2016inception}, MobileNet\cite{howard2017mobilenets}, VGG16\cite{simonyan2015very}, and ResNet50\cite{he2015deep}. This comparison aims to provide an overall assessment of the proposed model relative to existing baseline CNNs widely utilized for breast cancer classification. All models, including the proposed one, are trained utilizing the identical dataset, and the results are presented in Table \ref{tab:clsComp}. The performance of these models is evaluated based on the following metrics: Accuracy (Acc), Precision (Prec), Recall (Rec), and F1 Score (F1).

\begin{table}[h!]
 \caption{Performance Metrics Comparison of Proposed Classification Model with Other Baseline CNN Models}
  \centering
  \begin{tabular}{lcccccccc}
    \toprule
    &\multicolumn{4}{c}{Training Phase Metrics}& \multicolumn{4}{c}{Validation Phase Metrics} \\
    \cmidrule(r){2-5} \cmidrule(r){6-9}
    Model & Acc & Prec & Rec & F1 & Acc & Prec & Rec & F1 \\
    \midrule
    EfficientNetV2 & 0.926 & 0.931 & 0.920& 0.925& 0.871 & 0.871 & 0.871 & 0.871 \\
    MobileNetV2 & 0.935 & 0.948 & 0.925 & 0.936& 0.858 & 0.857 & 0.852 & 0.854 \\
    DenseNet121 & 0.928 & 0.938 & 0.925 & 0.931& 0.906 & 0.906 & 0.906 & 0.906 \\
    NASNetMobile & 0.942 & 0.947 & 0.941 & 0.944 & 0.911 & 0.913 & 0.904 & 0.908 \\
    Xception & 0.925 & 0.926 & 0.920& 0.923 & 0.917 & 0.917 & 0.917 & 0.917 \\
    InceptionV3 & 0.878& 0.897 & 0.862 & 0.879& 0.774 & 0.774 & 0.771 & 0.772 \\
    InceptionResNetV2 & 0.958 & 0.961 & 0.958 & 0.959& 0.947 & 0.957 & 0.901 & 0.928 \\
    MobileNet & 0.956 & 0.957 & 0.948 & 0.952& 0.872& 0.874& 0.871& 0.873\\
    VGG16 & 0.86 & 0.889 & 0.841 & 0.864& 0.861 & 0.877 & 0.803 & 0.838 \\
    ResNet50 & 0.837 & 0.866 & 0.805 & 0.834& 0.761 & 0.761 & 0.761 & 0.761 \\
    \textbf{CSFEC-Net (Proposed)} & \textbf{0.989} & \textbf{0.994} & \textbf{0.992} & \textbf{0.993} & \textbf{0.992} & \textbf{0.993} & \textbf{0.991} & \textbf{0.991} \\
    \bottomrule
  \end{tabular}
  \label{tab:clsComp}
\end{table}

From Table 5, it is evident that the proposed CSFEC-Net model has outperformed all baseline CNN models in terms of performance evaluation metrics. EfficientNetV2 overfits on the data due to difficulty in generalizing the nuanced features of breast cancer like irregular margins of malignant lesions or varying degrees of echogenicity observed in ultrasound images. MobileNetV2's lightweight architecture struggles with the detailed analysis required to detect early signs of breast cancer, such as subtle changes in echotexture or the presence of microcalcifications within lesions. While DenseNet121 benefits from dense connectivity for feature reuse, its performance in identifying specific breast cancer markers like the orientation and distribution of calcifications or the assessment of lesion vascularity is compromised. The lack of accurate irregular mass shapes and variation in posterior acoustic shadowing are some features that characterize breast cancer that might not be picked well by NASNetMobile because it was designed simply for mobile applications. On the other hand, Xception does not take advantage of the spatial dependencies that are important in determining certain indicators of breast cancer, such as the pattern of calcifications or the echogenicity of surrounding tissue.

InceptionV3's design compromise for computational efficiency limits its ability to analyze multidimensional data, characteristic of breast cancer ultrasound images, particularly in detecting subtle architectural distortions or changes in tissue echotexture. Despite its sophisticated architecture, InceptionResNetV2 does not optimally resonate with the need to identify very specific, disease-related features like the texture and margin irregularities of masses or the presence of ductal abnormalities. MobileNet's efficiency does dictate how its features should not be very deep, thus failing on the level of detail required during feature extraction by breast cancer ultrasound images. The simplicity and related shallowness of VGG16 struggle with a more intricate analysis needed for detecting and classifying features such as posterior acoustic enhancement, accounting for lower accuracy in validation tests. Features such as the assessment of lesion margins might not be adequately learned due to limitations in the ResNet50’s depth and focus. The proposed model's integrated CSFEM effectively leverages multi-level attention mechanisms for the identification of critical features such as calcifications, architectural distortions, and mass margins across diverse tumor segmentation qualities. These enhancements allow the model to capture the complex, heterogeneous pathology of breast cancer evident in ultrasound imagery. 

\begin{table}[h!]
 \caption{Performance Metrics Comparison of Proposed Classification Model with Other Models}
  \centering
  \begin{tabular}{p{4cm}cccc}
    \toprule
    \textbf{Classification Model} & \textbf{Accuracy (\%)} & \textbf{Precision (\%)} & \textbf{Recall (\%)} & \textbf{F1-Score (\%)} \\
    \midrule
    Fine Tuned VGG16 and Fine Tuned VGG19 ensemble model \cite{hameed2020breast} & 95.29 & 95.46 & 95.20& 95.29 \\
    CNN-based Ensemble Learner with MLP meta classifier \cite{das2021breast} & 98.08 & 98.41 & 98.82& 98.81\\
    BCCNN \cite{abunasser2023convolution} & 98.31& 98.39 & 98.30& 98.28 \\
    ResNet50 hybrid with SVM \cite{salama2020novel} & 97.98 & 96.51 & 97.63 & 95.97 \\
    Deep CNN with Fuzzy merging \cite{krithiga2020deep} & 98.62& 92.31& 94.70& 93.53\\
    Xception + SVM R \cite{sharma2022xception} & 96.25 & 96.12& 96.02& 96.01\\
    Grid-based deep feature generator + DNN classifier \cite{liu2022artificial} & 97.18 & 97.45 & 96.18 & 96.79 \\
    InceptionV3 with residual connections \cite{sirjani2023novel} & 91.03& 85.05& 96.01& 92.02\\
    EDLCDS-BCDC \cite{ragab2022ensemble} & 95.15 & 97.35 & 94.74 & 96.92 \\
    AlexNet, ResNet50 and MobileNetV2 Hybrid feature extractor + mRMR + SVM \cite{eroglu2021convolutional} & 95.60& 95.69 & 95.61 & 95.65 \\
    \textbf{CSFEC-Net (Proposed)}& \textbf{99.20}& \textbf{99.32}& \textbf{99.14}& \textbf{99.1}\\
    \bottomrule
  \end{tabular}
  \label{tab:clsComp2}
\end{table}

From the results presented in Table \ref{tab:clsComp2}, it is apparent that the CSFEC-Net model proposed in this research outperforms all other models in existing research. The assembly of Fine Tuned VGG16 and VGG19 \cite{hameed2020breast} achieves moderate performance with accuracy and F1-scores around 95\%. Its performance is relatively low, indicating potential limitations in its ability to capture the complexity of breast cancer pathology fully. CNN-based Ensemble Learner with MLP Meta Classifier \cite{das2021breast} has shown high performance with an accuracy of 98\% but has struggled with identifying subtle changes in the irregular shapes of masses. BCCNN \cite{abunasser2023convolution} shows promising results with metrics around 98\%. However, the slight variation in F1-score compared to the highest performers suggests it faces challenges in maintaining a balance between precision and recall, essential for minimizing errors in breast cancer diagnosis. 

The SVM hybrid with ResNet50 \cite{salama2020novel} shows high recall but is weak in precision. Thus, while this model is good at identifying the number of positive cases, it has difficulty discerning benign lesions from malignant ones, which tends to lead it to false positives. The precision score drops markedly in the case of Deep CNN with Fuzzy Merging \cite{krithiga2020deep}, alerting attention to a serious failure in its ability to classify breast cancer cases accurately. This implies that it might be good at identifying macro patterns but poor at capturing the important micro ones that make for accurate diagnosis. The Xception combined with SVM R \cite{sharma2022xception} shows a fair performance of around 96\% which points to some relative inefficiency as compared to the other models with respect to feature extraction capabilities, which makes it inefficient for realistic use. Grid-based Deep Feature Generator with DNN Classifier \cite{liu2022artificial} has high precision scores, but the insignificant differences in recall and F1-score might point toward some ineffectiveness in the full capture of relevant pathological features, which will reduce its overall efficacy.

InceptionV3 with Residual Connections achieves a recall of very high values while at the same time having a precision value substantially lower than the recall, indicating a huge asymmetry in diagnostic capabilities. Such imbalances are usually due to very great discrimination difficulties in benign cases against malignant ones, which is necessary to cut down false positives. On the other hand, EDLCDS-BCDC gives a moderate report in that range of metrics, between 95\% to 97\%, which suggests it cannot be very accurate on spotting very minute differences. AlexNet, ResNet50, and MobileNetV2 Hybrid Feature Extractor with mRMR and SVM shows outstanding performance with an accuracy and F1-score of about 95\%. This limitation of the model suggests that it is unable to fully adapt to the complex and diverse pathology of breast cancer and can be improved.

The proposed CSFEC-Net model has also outperformed all other models in this comparison with high performance in almost every metric evaluated. The key factor for this performance is its rigorous architectural design, inclusive of an exclusive CSFEM with multiscale attentions, to enable the most precise and optimal extraction of the subtle pathological features of breast cancer from the data. This individualized strategy provides both high accuracy and tremendous precision and recall and demonstrates robustness and reliability in clinical applications for breast cancer classification.

\section{Conclusion and Future Direction}

In this paper, we propose two novel DL frameworks to overcome the two important hurdles in the diagnosis of breast cancer using ultrasound imaging: PMAD-LinkNet for segmentation and CSFEC-Net for classification. The segmentation framework leverages the Precision Mapping Mechanism (PMM) for locations to be spatially mapped automatically, thus allowing accurate tumor boundary refinement in the presence of morphological variations and disparate image quality. The classification framework features the Component-Specific Feature Enhancement Module (CSFEM), which utilizes multi-level attention to emphasize discriminative features between benign, malignant, and normal tissues.

While the suggested frameworks have demonstrated remarkable results, it does not stop the late work possible in this area. The generalizability of the models could be improved by collecting larger samples with greater diversity and by augmenting data multimodally. For instance, incorporate images of mammograms and obtain magnetic resonance imaging (MRI) data. Second, using explainable AI techniques could go a long way toward improving the interpretability of model decisions and hence increasing trust and clinical adoption. Third, real-time implementations and optimizations should be researched and developed for deployment in edge devices so that it could enable more wide accessibility in non-resource settings. Finally, adapting the frameworks to all other possible imaging modalities and applications in medicine would further show their versatility and impact on healthcare. In doing so, we will make those avenues possible that will enhance the applicability and effectiveness of proposed methods toward improving early detection and treatment outcomes concerning breast cancer and many more critical diseases.

\bibliographystyle{IEEEtran}
\bibliography{manuscript.bib}

\begin{thebibliography}{10}
\providecommand{\url}[1]{#1}
\csname url@samestyle\endcsname
\providecommand{\newblock}{\relax}
\providecommand{\bibinfo}[2]{#2}
\providecommand{\BIBentrySTDinterwordspacing}{\spaceskip=0pt\relax}
\providecommand{\BIBentryALTinterwordstretchfactor}{4}
\providecommand{\BIBentryALTinterwordspacing}{\spaceskip=\fontdimen2\font plus
\BIBentryALTinterwordstretchfactor\fontdimen3\font minus \fontdimen4\font\relax}
\providecommand{\BIBforeignlanguage}[2]{{%
\expandafter\ifx\csname l@#1\endcsname\relax
\typeout{** WARNING: IEEEtran.bst: No hyphenation pattern has been}%
\typeout{** loaded for the language `#1'. Using the pattern for}%
\typeout{** the default language instead.}%
\else
\language=\csname l@#1\endcsname
\fi
#2}}
\providecommand{\BIBdecl}{\relax}
\BIBdecl

\bibitem{stewart2014world}
B.~Stewart and C.~Wild, \emph{World Cancer Report 2014}.\hskip 1em plus 0.5em minus 0.4em\relax Geneva, Switzerland: WHO Press, 2014.

\bibitem{whoBreastCancer}
{World Health Organization}, ``Breast cancer,'' \url{http://www.who.int/cancer/prevention/diagnosis-screening/breast-cancer/en/}, accessed: 2024-07-03.

\bibitem{sun2017risk}
Y.~Sun, Z.~Zhao, Z.~Yang, F.~Xu, H.~Lu, Z.~Zhu, W.~Shi, J.~Jiang, P.~Yao, and H.~Zhu, ``Risk factors and preventions of breast cancer,'' \emph{Int J Biol Sci}, Nov 2017.

\bibitem{tarique2015fourier}
M.~Tarique, F.~Elzahra, A.~Hateem, and M.~Mohammad, ``Fourier transform based early detection of breast cancer by mammogram image processing,'' \emph{J Biomed Eng Med Imaging}, vol.~24, p.~17, 2015.

\bibitem{acsDiagnosis}
{American Cancer Society}, ``How is breast cancer diagnosed?'' \url{http://www.cancer.org/cancer/breastcancer/detailedguide/breast-cancer-diagnosis}, 2014, accessed: September 20, 2017.

\bibitem{sadoughi2018artificial}
F.~Sadoughi, Z.~Kazemy, F.~Hamedan, L.~Owji, M.~Rahmanikatigari, and T.~Azadboni, ``Artificial intelligence methods for the diagnosis of breast cancer by image processing: a review,'' \emph{Breast Cancer (Dove Med Press)}, vol.~10, pp. 219--230, Nov 2018.

\bibitem{benson2009early}
J.~Benson, I.~Jatoi, M.~Keisch, F.~Esteva, A.~Makris, and V.~Jordan, ``Early breast cancer,'' \emph{Lancet}, vol. 373, no. 9673, pp. 1463--79, Apr 2009.

\bibitem{alz}
\BIBentryALTinterwordspacing
P.~V, S.~Venkatraman, A.~A, P.~K. S, A.~S. A, and K.~A, ``Leveraging bi-focal perspectives and granular feature integration for accurate reliable early alzheimer's detection,'' 2024. [Online]. Available: \url{https://arxiv.org/abs/2407.10921}
\BIBentrySTDinterwordspacing

\bibitem{litjens2017survey}
G.~Litjens, T.~Kooi, B.~Bejnordi, A.~Setio, F.~Ciompi, M.~Ghafoorian, J.~van~der Laak, B.~van Ginneken, and C.~S{\'a}nchez, ``A survey on deep learning in medical image analysis,'' \emph{Med Image Anal}, vol.~42, pp. 60--88, Dec 2017.

\bibitem{rashed2019deep}
E.~Rashed and M.~El~Seoud, ``Deep learning approach for breast cancer diagnosis,'' in \emph{Proceedings of the 8th International Conference on Software and Information Engineering}, Apr 2019, pp. 243--247.

\bibitem{ramesh2022segmentation}
S.~Ramesh, S.~Sasikala, S.~Gomathi \emph{et~al.}, ``Segmentation and classification of breast cancer using novel deep learning architecture,'' \emph{Neural Comput \& Applic}, vol.~34, pp. 16\,533--16\,545, 2022.

\bibitem{skin}
S.~S, S.~Venkatraman, S.~Malarvannan, S.~A, and S.~S, ``A comprehensive study on skin cancer detection using deep learning,'' in \emph{2024 IEEE 3rd World Conference on Applied Intelligence and Computing (AIC)}, 2024, pp. 749--754.

\bibitem{osareh2010machine}
A.~Osareh and B.~Shadgar, ``Machine learning techniques to diagnose breast cancer,'' in \emph{2010 5th International Symposium on Health Informatics and Bioinformatics}, Ankara, Turkey, 2010, pp. 114--120.

\bibitem{li2019classification}
Y.~Li, J.~Wu, and Q.~Wu, ``Classification of breast cancer histology images using multi-size and discriminative patches based on deep learning,'' \emph{IEEE Access}, vol.~7, pp. 21\,400--21\,408, 2019.

\bibitem{zheng2020deep}
J.~Zheng, D.~Lin, Z.~Gao, S.~Wang, M.~He, and J.~Fan, ``Deep learning assisted efficient adaboost algorithm for breast cancer detection and early diagnosis,'' \emph{IEEE Access}, 2020.

\bibitem{lotter2021robust}
W.~Lotter, A.~Diab, B.~Haslam \emph{et~al.}, ``Robust breast cancer detection in mammography and digital breast tomosynthesis using an annotation-efficient deep learning approach,'' \emph{Nat Med}, vol.~27, pp. 244--249, 2021.

\bibitem{saber2021novel}
A.~Saber, M.~Sakr, O.~Abo-Seida, A.~Keshk, and H.~Chen, ``A novel deep-learning model for automatic detection and classification of breast cancer using the transfer-learning technique,'' \emph{IEEE Access}, vol.~9, pp. 71\,194--71\,209, 2021.

\bibitem{cho2022deep}
S.~Cho, N.~Baek, and K.~Park, ``Deep learning-based multi-stage segmentation method using ultrasound images for breast cancer diagnosis,'' \emph{J King Saud Univ - Comput Inf Sci}, vol.~34, no.~10, pp. 10\,273--10\,292, 2022.

\bibitem{wang2016deep}
D.~Wang, A.~Khosravi, R.~Gargeya, H.~Irshad, and A.~Beck, ``Deep learning for identifying metastatic breast cancer,'' \emph{arXiv preprint arXiv:1606.05718}, 2016.

\bibitem{hamed2020deep}
G.~Hamed, T.~Helmy, H.~Badawi, and M.~Shawky, ``Deep learning in breast cancer detection and classification,'' in \emph{Advances in Intelligent Systems and Computing}, 2020.

\bibitem{balkenende2022application}
L.~Balkenende, J.~Teuwen, and R.~Mann, ``Application of deep learning in breast cancer imaging,'' \emph{Semin Nucl Med}, vol.~52, no.~5, 2022.

\bibitem{shen2019deep}
L.~Shen, L.~Margolies, J.~Rothstein \emph{et~al.}, ``Deep learning to improve breast cancer detection on screening mammography,'' \emph{Sci Rep}, vol.~9, p. 12495, 2019.

\bibitem{han2017breast}
Z.~Han, B.~Wei, Y.~Zheng \emph{et~al.}, ``Breast cancer multi-classification from histopathological images with structured deep learning model,'' \emph{Sci Rep}, vol.~7, p. 4172, 2017.

\bibitem{sizilio2012fuzzy}
G.~Sizilio, C.~Leite, A.~Guerreiro, and A.~Neto, ``Fuzzy method for pre-diagnosis of breast cancer from the fine needle aspirate analysis,'' \emph{Biomed Eng Online}, vol.~11, no.~1, p.~83, 2012.

\bibitem{sarkar2000application}
M.~Sarkar and T.~Leong, ``Application of k-nearest neighbours algorithm on breast cancer diagnosis problem,'' in \emph{Proceedings of the AMIA Symposium}.\hskip 1em plus 0.5em minus 0.4em\relax American Medical Informatics Association, 2000, pp. 793--797.

\bibitem{foster2014machine}
K.~Foster, R.~Koprowski, and J.~Skufca, ``Machine learning, medical diagnosis, and biomedical engineering research-commentary,'' \emph{Biomed Eng Online}, vol.~13, no.~1, p.~94, 2014.

\bibitem{wei2009microcalcification}
L.~Wei, Y.~Yang, and R.~Nishikawa, ``Microcalcification classification assisted by content-based image retrieval for breast cancer diagnosis,'' \emph{Pattern Recognition}, vol.~42, no.~6, pp. 1126--1132, 2009.

\bibitem{kaggle_breast_ultrasound}
\BIBentryALTinterwordspacing
A.~Shah, ``Breast ultrasound images dataset,'' 2020, accessed: 2024-07-03. [Online]. Available: \url{https://www.kaggle.com/datasets/aryashah2k/breast-ultrasound-images-dataset}
\BIBentrySTDinterwordspacing

\bibitem{chaurasia2017linknet}
A.~Chaurasia and E.~Culurciello, ``Linknet: Exploiting encoder representations for efficient semantic segmentation,'' in \emph{2017 IEEE Visual Communications and Image Processing (VCIP)}.\hskip 1em plus 0.5em minus 0.4em\relax IEEE, 2017, pp. 1--4.

\bibitem{szegedy2016inception}
\BIBentryALTinterwordspacing
C.~Szegedy, S.~Ioffe, and V.~Vanhoucke, ``Inception-v4, inception-resnet and the impact of residual connections on learning,'' \emph{CoRR}, vol. abs/1602.07261, 2016. [Online]. Available: \url{http://arxiv.org/abs/1602.07261}
\BIBentrySTDinterwordspacing

\bibitem{selvaraju2017gradcam}
R.~Selvaraju, M.~Cogswell, A.~Das, R.~Vedantam, D.~Parikh, and D.~Batra, ``Grad-cam: Visual explanations from deep networks via gradient-based localization,'' in \emph{Proceedings of the IEEE International Conference on Computer Vision}, 2017, pp. 618--626.

\bibitem{almajalid2018development}
R.~Almajalid, J.~Shan, Y.~Du, and M.~Zhang, ``Development of a deep-learning-based method for breast ultrasound image segmentation,'' in \emph{2018 17th IEEE International Conference on Machine Learning and Applications (ICMLA)}.\hskip 1em plus 0.5em minus 0.4em\relax IEEE, 2018, pp. 1103--1108.

\bibitem{yue2022deep}
W.~Yue, H.~Zhang, J.~Zhou \emph{et~al.}, ``Deep learning-based automatic segmentation for size and volumetric measurement of breast cancer on magnetic resonance imaging,'' \emph{Front Oncol}, vol.~12, p. 984626, 2022.

\bibitem{zhang2023fully}
S.~Zhang, M.~Liao, J.~Wang \emph{et~al.}, ``Fully automatic tumor segmentation of breast ultrasound images with deep learning,'' \emph{J Appl Clin Med Phys}, vol.~24, p. e13863, 2023.

\bibitem{li2021multi}
J.~Li, L.~Cheng, T.~Xia, H.~Ni, and J.~Li, ``Multi-scale fusion u-net for the segmentation of breast lesions,'' \emph{IEEE Access}, vol.~9, pp. 137\,125--137\,139, 2021.

\bibitem{tan2021efficientnetv2}
\BIBentryALTinterwordspacing
M.~Tan and Q.~Le, ``Efficientnetv2: Smaller models and faster training,'' \emph{CoRR}, vol. abs/2104.00298, 2021. [Online]. Available: \url{https://arxiv.org/abs/2104.00298}
\BIBentrySTDinterwordspacing

\bibitem{sandler2018inverted}
\BIBentryALTinterwordspacing
M.~Sandler, A.~Howard, M.~Zhu, A.~Zhmoginov, and L.~Chen, ``Inverted residuals and linear bottlenecks: Mobile networks for classification, detection and segmentation,'' \emph{CoRR}, vol. abs/1801.04381, 2018. [Online]. Available: \url{http://arxiv.org/abs/1801.04381}
\BIBentrySTDinterwordspacing

\bibitem{huang2016densely}
\BIBentryALTinterwordspacing
G.~Huang, Z.~Liu, and K.~Weinberger, ``Densely connected convolutional networks,'' \emph{CoRR}, vol. abs/1608.06993, 2016. [Online]. Available: \url{http://arxiv.org/abs/1608.06993}
\BIBentrySTDinterwordspacing

\bibitem{zoph2017learning}
\BIBentryALTinterwordspacing
B.~Zoph, V.~Vasudevan, J.~Shlens, and Q.~Le, ``Learning transferable architectures for scalable image recognition,'' \emph{CoRR}, vol. abs/1707.07012, 2017. [Online]. Available: \url{http://arxiv.org/abs/1707.07012}
\BIBentrySTDinterwordspacing

\bibitem{chollet2016xception}
\BIBentryALTinterwordspacing
F.~Chollet, ``Xception: Deep learning with depthwise separable convolutions,'' \emph{CoRR}, vol. abs/1610.02357, 2016. [Online]. Available: \url{http://arxiv.org/abs/1610.02357}
\BIBentrySTDinterwordspacing

\bibitem{szegedy2015rethinking}
\BIBentryALTinterwordspacing
C.~Szegedy, V.~Vanhoucke, S.~Ioffe, J.~Shlens, and Z.~Wojna, ``Rethinking the inception architecture for computer vision,'' \emph{CoRR}, vol. abs/1512.00567, 2015. [Online]. Available: \url{http://arxiv.org/abs/1512.00567}
\BIBentrySTDinterwordspacing

\bibitem{howard2017mobilenets}
\BIBentryALTinterwordspacing
A.~Howard, M.~Zhu, B.~Chen, D.~Kalenichenko, W.~Wang, T.~Weyand \emph{et~al.}, ``Mobilenets: Efficient convolutional neural networks for mobile vision applications,'' \emph{CoRR}, vol. abs/1704.04861, 2017. [Online]. Available: \url{http://arxiv.org/abs/1704.04861}
\BIBentrySTDinterwordspacing

\bibitem{simonyan2015very}
\BIBentryALTinterwordspacing
K.~Simonyan and A.~Zisserman, ``Very deep convolutional networks for large-scale image recognition,'' \emph{arXiv [Cs.CV]}, vol. abs/1409.1556, 2015. [Online]. Available: \url{http://arxiv.org/abs/1409.1556}
\BIBentrySTDinterwordspacing

\bibitem{he2015deep}
\BIBentryALTinterwordspacing
K.~He, X.~Zhang, S.~Ren, and J.~Sun, ``Deep residual learning for image recognition,'' \emph{CoRR}, vol. abs/1512.03385, 2015. [Online]. Available: \url{http://arxiv.org/abs/1512.03385}
\BIBentrySTDinterwordspacing

\bibitem{hameed2020breast}
Z.~Hameed, S.~Zahia, B.~Garcia-Zapirain, J.~Javier~Aguirre, and A.~María~Vanegas, ``Breast cancer histopathology image classification using an ensemble of deep learning models,'' \emph{Sensors}, vol.~20, no.~16, p. 4373, 2020.

\bibitem{das2021breast}
A.~Das, M.~Mohanty, P.~Mallick, P.~Tiwari, K.~Muhammad, and H.~Zhu, ``Breast cancer detection using an ensemble deep learning method,'' \emph{Biomed Signal Process Control}, vol.~70, p. 103009, 2021.

\bibitem{abunasser2023convolution}
B.~Abunasser, M.~Al-Hiealy, I.~Zaqout, and S.~Abu-Naser, ``Convolution neural network for breast cancer detection and classification using deep learning,'' \emph{Asian Pac J Cancer Prev}, vol.~24, no.~2, pp. 531--544, 2023.

\bibitem{salama2020novel}
W.~Salama, A.~Elbagoury, and M.~Aly, ``Novel breast cancer classification framework based on deep learning,'' \emph{IET Image Process}, vol.~14, no.~13, pp. 3254--3259, 2020.

\bibitem{krithiga2020deep}
R.~Krithiga and P.~Geetha, ``Deep learning based breast cancer detection and classification using fuzzy merging techniques,'' \emph{Mach Vis Appl}, vol.~31, p.~63, 2020.

\bibitem{sharma2022xception}
S.~Sharma and S.~Kumar, ``The xception model: A potential feature extractor in breast cancer histology images classification,'' \emph{ICT Express}, vol.~8, no.~1, pp. 101--108, 2022.

\bibitem{liu2022artificial}
H.~Liu, G.~Cui, Y.~Luo, Y.~Guo, L.~Zhao, Y.~Wang \emph{et~al.}, ``Artificial intelligence-based breast cancer diagnosis using ultrasound images and grid-based deep feature generator,'' \emph{Int J Gen Med}, vol.~15, pp. 2271--2282, 2022.

\bibitem{sirjani2023novel}
N.~Sirjani, M.~Ghelich~Oghli, M.~Tarzamni, M.~Gity, A.~Shabanzadeh, P.~Ghaderi \emph{et~al.}, ``A novel deep learning model for breast lesion classification using ultrasound images: A multicenter data evaluation,'' \emph{Phys Medica}, vol. 107, p. 102560, 2023.

\bibitem{ragab2022ensemble}
M.~Ragab, A.~Albukhari, J.~Alyami, and R.~Mansour, ``Ensemble deep-learning-enabled clinical decision support system for breast cancer diagnosis and classification on ultrasound images,'' \emph{Biology}, vol.~11, p. 439, 2022.

\bibitem{eroglu2021convolutional}
Y.~Eroğlu, M.~Yildirim, and A.~Çinar, ``Convolutional neural networks based classification of breast ultrasonography images by hybrid method with respect to benign, malignant, and normal using mrmr,'' \emph{Comput Biol Med}, vol. 133, p. 104407, 2021.

\end{thebibliography}

\end{document}